\documentclass[aps,pre,showpacs,twocolumn]{revtex4}
\usepackage{mathrsfs}
\usepackage{amssymb}
\usepackage{amsmath}
\usepackage{graphicx}
\usepackage{dcolumn}
\usepackage{bm}

\begin{document}

\title{Quantum distance and the Euler number index of the Bloch band in a one-dimensional
spin model}
\author{Yu-Quan Ma}
\email{mayuquan@iphy.ac.cn}
\affiliation{School of Applied Science, Beijing Information Science and Technology
University, Beijing 100192, China}
\date{\today }

\begin{abstract}
We study the Riemannian metric and the Euler characteristic number of the
Bloch band in a one-dimensional spin model with multi-site spins exchange interactions.
The Euler number of the Bloch band originates from the Gauss-Bonnet theorem
on the topological characterization of the closed Bloch states manifold in
the first Brillouin zone. We study this approach analytically in a
transverse field XY spin chain with three-site spin coupled interactions. We
define a class of cyclic quantum distance on the Bloch band and on the
ground state, respectively, as a local characterization for quantum phase
transitions. Specifically, we give a general formula for the Euler number by
means of the Berry curvature in the case of two-band models, which reveals
its essential relation to the first Chern number of the band insulators.
Finally, we show that the ferromagnetic-paramagnetic phases transition in
zero-temperature can be distinguished by the Euler number of the Bloch band.
\end{abstract}

\pacs{03.65.Vf, 73.43.Nq, 75.10.Pq, 05.70.Jk}
\maketitle



\section{Introduction}

The topological nature of quantum states has become a key ingredient in
understanding the novel quantum phases of condensed-matter systems in low
temperatures. Since the discovery of the Berry phase as a geometric phase
picked up from the cyclic adiabatic evolutions of the Hamiltonian eigenstate
and its holonomy interpretation on the $U(1)$ line bundle with parallel
transport, many important findings on the topological nature of the quantum
matter have come into physics, i.e., the quantized Hall conductance \cite%
{Laughlin,TKNN,Niu1984}, adiabatic pumping \cite{Thouless,Niu}, topological
insulators and superconductivity \cite{Kane,Fu07,Ryu08,Hasan,Qi}, and
recently the fractional Chern insulators in flat bands \cite%
{Tang,Sun,Neupert1}.

In recent years, lots of attention has been attracted into understanding the
quantum phase transitions (QPTs) \cite{Sachdev,Sondhi,Vojta} from the
quantum information and the Hilbert space geometry aspects \cite%
{Bengtsson,Ortiz}. Essentially, a QPT is the result of the competing
ground-state phases driven by the quantum fluctuations, which can be
witnessed by some qualitative changes of the ground-state properties, i.e.,
quantum entanglement \cite{ent1,ent2,ent3,ent4,ent5}, entanglement entropy
\cite{ee1,ee2}, quantum discord \cite{qd1,qd2,qd3}, quantum fidelity and the
fidelity susceptibility \cite%
{Zanardi,Gu07,Gu10,Chen,Yang,Zhao,Damski1,Dutta,Damski2,Nishiyama}, the Berry
phase \cite%
{Berry,Simon,Resta94,Xiao,Pachos,Hamma,Zhu,Ma2009,Hatsugai,Fufu,Ma2012}, and
the quantum geometric tensor \cite%
{Provost,Berry1989,Resta,Venuti,Zanardi07,Ma2010,Ryu,Haldane,Neupert2}.

The ground-state geometric tensor, as an intrinsic metric on the
ground-state complex manifold, is naturally expected to shed some light on
the geometric characterization of QPTs. Mathematically, the quantum geometric
tensor, also called the Fubini-Study metric, is a Hermitian metric on the
complex projective space of the quantum states. Physically, the
(non-Abelian) geometric tensor originates from defining a local $U(n)$ gauge
invariant quantum distance between two states in a parameterized Hilbert
space \cite{Ma2010}. The quantum geometric tensor brings a Riemannian structure
to the parameterized quantum states, where the corresponding Riemannian
metric is given by the real part of the geometric tensor. Meanwhile, its
imaginary part was later found to be just the Berry curvature (up to a
constant coefficient). Specifically, the ground-state geometric tensor
provides a unified mechanism from the aspect of information-geometry to
understand the critical behaviors in quantum many-body systems.

Recently, a direct measurement of the Zak phase \cite{Zak}, as a Berry phase
of a 1D Bloch band, has been achieved in one-dimensional (1D) optical lattices \cite{Bloch}.
For the geometric tensor of the Bloch band, some interesting measurable
consequences have been proposed by relating the geometric tensor of band
insulators to the current noise spectrum \cite{Neupert2}. A more interesting
question is whether there exists some topological characterization related
to the Riemannian metric of the Bloch bands? Very recently, a topological
Euler number of the Bloch band was proposed to distinguish nontrivial
topological phases in gapped free fermionic systems. This fact was pointed
out in our previous work \cite{Ma2013} and later by Kolodrubetz et. al \cite%
{Kolodrubetz}.

In this work, we study the local and topological properties of the Bloch
band in a 1D transverse field XY spin-1/2 model with three-site spin
interactions. The system exhibits a nonzero transverse magnetization at the
zero transverse field due to its multi-site spins exchange interactions. In
order to obtain a well-defined geometric tensor in the crystal momentum
space, we introduce an extra 1D parameter space by subjecting the spin
system to a local gauge transformation, which in fact puts the Hamiltonian
of the system on a torus $T^{2}$ in a 1+1D crystal momentum space without
changing its energy spectrum. By using of the quantum Riemannian metric on
the Bloch states manifold, we introduce a class of cyclic quantum distance
as a local characterization for quantum phase transitions. Particularly, we
derive the Euler characteristic number of the Bloch band analytically via
the Gauss-Bonnet theorem on the Bloch states manifold in the first Brillouin
zone. A general formula for the Euler number is obtained by means of the
Berry curvature in the case of two-band models, which also reveals its
relation to the first Chern number of the band insulators. Finally, we show
that the ferromagnetic and paramagnetic quantum phase transitions can be
distinguished by the different Euler numbers of the Bloch band.

\section{The model}

We consider a 1D anisotropic XY spin-1/2 model with three-site spin exchange
interactions in a transverse field. This spin model exhibits a nonzero
transverse magnetization at the zero transverse field due to its multiple
sites spin coupling and shows a rich ground-state phase diagram \cite%
{Lou,Zvyagin,Perk,Cheng}. The Hamiltonian reads
\begin{eqnarray}
H_{\text{S}} &=&\sum_{l\in N}^{\text{PBC}}-\left( 1+\gamma \right)
S_{l}^{x}S_{l+1}^{x}-\left( 1-\gamma \right) S_{l}^{y}S_{l+1}^{y}  \notag \\
&&-2\delta \left(
S_{l-1}^{x}S_{l}^{z}S_{l+1}^{x}+S_{l-1}^{y}S_{l}^{z}S_{l+1}^{y}\right)
-hS_{l}^{z},
\end{eqnarray}%
where $S_{l}^{\alpha }$ ($\alpha =x,y,z;l\in N$) is the Pauli operator on
the local site $l$, $N$ denotes the total number of the sites, $\gamma $ is the anisotropy parameter in the in-plane
interaction, $\delta $ denotes the three-site XZX+YZY type spins exchange
interactions, $h$ is the transverse magnetic field, and the periodic
boundary condition (PBC) has been imposed on this model.

Here we will show that the quantum critical points of the system can be
witnessed by some local geometric characterization, i.e., the Riemannian
metric on the Bloch band, and some partial derivative of the ground-state
quantum distance. Particularly, we show that the zero-temperature phases
diagram of the system can be marked by a nontrivial topological Euler number
index of the Bloch band in the crystal momentum space.

In order to investigate the ground-state geometric tensor for the system, we
need to define the metric tensor on a 2D parameter space. This can be
achieved by subjecting the system to a local gauge transformation $H_{\text{S%
}}(\varphi )=g(\varphi )H_{\text{S}}g(\varphi )^{\dagger }$ by a twist
operator $g(\varphi )=\prod_{l}e^{i\varphi S_{l}^{z}}$, which makes the
system a rotation on the spin along the $z$-direction. It can be verified
that $H_{\text{S}}(\varphi )$ is $\pi $ periodic in $\varphi $ because the
quadratic form about the $x$ and $y$ axes appears symmetric in the
Hamiltonian. Considering the unitarity of the twist operator $g(\varphi )$,
the critical behavior and energy spectrum of the system are obviously
parameter $\varphi $ independent.

The spin Hamiltonian $H_{\text{S}}(\varphi )$ can be mapped exactly on a
spinless fermion Hamiltonian $H_{\text{F}}(\varphi )$ by the Jordan-Wigner
transformation $a_{l}=\prod_{m=1}^{l-1}\left( -2S_{m}^{z}\right) S_{l}^{-}$,
$a_{l}^{\dagger }=\prod_{m=1}^{l-1}\left( -2S_{m}^{z}\right) S_{l}^{+}$,
where $S_{l}^{\pm }=S_{l}^{x}\pm iS_{l}^{x}$ denote the spin ladder
operators and $a_{l}$, $a_{l}^{\dagger }$ are the corresponding Fermion
annihilation and creation operators, respectively, on the local site $l$. After applying a
Fourier transformation $a_{l}=\frac{1}{\sqrt{N}}\sum_{k\in \text{Bz}%
}e^{ikl}c_{k}$, we can rewrite the fermion Hamiltonian as
\begin{equation}
H_{\text{F}}(\varphi )=\sum_{k\in \text{Bz}}\Psi _{k,\varphi }^{\dagger
}\left( \sum_{\alpha =1}^{3}d_{\alpha }\left( k,\varphi \right) \sigma
^{\alpha }\right) \Psi _{k,\varphi },  \label{ferham}
\end{equation}%
where $d_{1}\left( k,\varphi \right) =\frac{1}{2}\gamma \sin k\sin 2\varphi $%
, $d_{2}\left( k,\varphi \right) =\frac{1}{2}\gamma \sin k\cos 2\varphi $, $%
d_{3}\left( k,\varphi \right) =\frac{1}{2}\left( -h+\delta \cos 2k-\cos
k\right) $, $\Psi _{k,\varphi }^{\dagger }:=\left( c_{k}^{\dagger }\text{, }%
c_{-k}\right) $ and $\sigma ^{\alpha }$ denotes the the Pauli matrices,
represent the pseudo-spin degree of freedom.

The Bloch wave function can be expressed as
\begin{equation}
u_{\pm }\left( k,\varphi \right) =\frac{1}{\sqrt{2d\left( d\mp d_{3}\left(
k,\varphi \right) \right) }}\left(
\begin{array}{c}
d_{1}\left( k,\varphi \right) -id_{2}\left( k,\varphi \right) \\
\pm d-d_{3}\left( k,\varphi \right)%
\end{array}%
\right) ,  \label{Bloch wave}
\end{equation}%
and the corresponding energy spectrum is $E_{\pm }(k)=\pm d$, where $d:=%
\sqrt{\sum_{\alpha =1}^{3}d_{\alpha }^{2}\left( k,\varphi \right) }$. The
Hamiltonian can be diagonalized as $H(\varphi )=\sum_{k\in \text{Bz}%
}E_{+}(k)\alpha _{k,\varphi }^{\dagger }\alpha _{k,\varphi }+E_{-}(k)\beta
_{k,\varphi }^{\dagger }\beta _{k,\varphi }$, and the $\varphi $
parameterized ground-state $\left\vert e\left( \varphi \right)
\right\rangle $ is the filled fermion sea
\begin{equation}
\left\vert e\left( \varphi \right) \right\rangle =\prod_{k>0}\beta
_{-k,\varphi }^{\dagger }\beta _{k,\varphi }^{\dagger }\left\vert
0\right\rangle ,  \label{gs}
\end{equation}%
where the quasi-particle operators $\alpha _{k,\varphi }=\left[ u\left(
\varphi ,k\right) _{+}\right] ^{\dagger }\Psi _{k,\varphi }$ and $\beta
_{k,\varphi }=\left[ u\left( \varphi ,k\right) _{-}\right] ^{\dagger }\Psi
_{k,\varphi }$. Note that the Bloch Hamiltonian $\mathcal{H}(k,\varphi
):=\sum_{\alpha =1}^{3}d_{\alpha }\left( k,\varphi \right) \sigma ^{\alpha }$
is period $\pi $ on the parameter $\varphi $, that is $\mathcal{H}(k,0)=%
\mathcal{H}(k,\pi )$. On the other hand, the Bloch Hamiltonian $\mathcal{H}%
(k,\varphi )$ can be regarded periodic in the Brillouin zone up to a gauge
transformation $\mathcal{H}(k+G,\varphi )=e^{-iG\cdot r}\mathcal{H}%
(k,\varphi )e^{iG\cdot r}$, where $G$, and $r$ are the reciprocal lattice vector
and position vector, respectively. Note that in a lattice model, here the
gauge factor is just identically equal to $1$, and we have $\mathcal{H}%
(k+G,\varphi )=\mathcal{H}(k,\varphi )$. Hence, the Bloch Hamiltonian $%
\mathcal{H}(k,\varphi )$ has been put on a torus $T^{2}$ in a 1+1D crystal
momentum space.

\section{Geometric tensor on the Bloch states manifold}

To begin with, we give a brief discussion on the quantum geometric tensor of
the Bloch band. The quantum geometric tensor of the Bloch band can be
derived naturally from a gauge invariant distance between two Bloch states
on the $U(1)$ line bundle induced by the quantum adiabatic evolution of the
Bloch state $\left\vert {u}_{n}(k)\right\rangle $ of the $n$-th filled band.
The gauge invariant quantum distance between two states $\left\vert {u}_{n}{%
\left( k{+\delta k}\right) }\right\rangle $ {and }$\left\vert {u}_{n}{\left(
k\right) }\right\rangle $ is given by
\begin{equation}
dS^{2}=\sum_{\mu ,\upsilon }\langle {\partial _{\mu }u}_{n}|\left[
\boldsymbol{1-}\mathcal{P}_{n}\right] \left\vert {\partial _{{\nu }}u}%
_{n}\right\rangle {{dk^{\mu }dk^{\upsilon }}},  \label{ds}
\end{equation}%
where $\mathcal{P}_{n}=\left\vert {u}_{n}\right\rangle \left\langle {u}%
_{n}\right\vert $ is the projection operator, and $\mu ,\nu $ denote the
components $k^{\mu }$ and $k^{\nu }$, respectively. The quantum geometric
tensor is given by%
\begin{equation}
Q_{\mu {\nu }}=\langle {\partial _{\mu }u}_{n}|\left[ \boldsymbol{1-}%
\mathcal{P}_{n}\right] \left\vert {\partial _{{\nu }}u}_{n}\right\rangle .
\label{qgt}
\end{equation}

The underlying mechanism for the quantum distance can be understood as
follows: The term $\left\vert {\partial _{\mu }u}_{n}\right\rangle $ can be
decomposed in the complete Hilbert space as $\left\vert {\partial _{\mu }u}%
_{n}\right\rangle =\left\vert {D_{\mu }u}_{n}\right\rangle {+}\left[
\boldsymbol{1}-\mathcal{P}_{n}\right] {{\left\vert {\partial _{\mu }u}%
_{n}\right\rangle }}$, where $\left\vert {D_{\mu }u}_{n}\right\rangle =%
\mathcal{P}_{n}\left\vert {\partial _{\mu }u}_{n}\right\rangle $ is the
covariant derivative of $\left\vert {u}_{n}\right\rangle $ on the line
bundle. Under the condition of the quantum adiabatic evolution, the
evolution of $\left\vert {u}_{n}{\left( k\right) }\right\rangle $ to $%
\left\vert {u}_{n}{\left( k+\delta k\right) }\right\rangle $ will undergo a
parallel transport, that is $\left\vert {D_{\mu }u}_{n}{\left( k\right) }%
\right\rangle =0$, which will lead to a gauge invariant quantum distance as
Eq. (\ref{ds}). The geometric tensor Eq. (\ref{qgt}) can be rewritten as $%
Q_{\mu {\nu }}=\mathcal{G}_{\mu {\nu }}-i\mathcal{F}_{\mu {\nu }}/2$, where $%
\mathcal{G}_{\mu {\nu }}:=$Re$Q_{\mu {\nu }}\ $can be verified as a
Riemannian metric, which establishes a Riemannian manifold of the Bloch
states. It can be verified that the quantum distance is only depend on the
real part of the quantum geometric tensor, that is $dS^{2}=\sum_{\mu
,\upsilon }\mathcal{G}_{\mu {\nu }}{{dk^{\mu }dk^{\upsilon }}}${,} because
the term{\ }$\mathcal{F}_{\mu {\nu }}:=-2$\text{Im}$Q_{\mu {\nu }}$ is
canceled out in the summation of the distance due to its antisymmetry.
However, the term{\ }$\mathcal{F}_{\mu {\nu }}$ can be associated to a two-form $\mathcal{F}=\sum_{\mu ,\upsilon }\mathcal{F}_{\mu {\nu }}{dk^{\mu
}\wedge dk^{\nu }}$, which is nothing but the Berry curvature.

\subsection{Riemannian metric and the cyclic quantum distance of the Bloch
band}

The Riemannian metric of the Bloch band is given by $\mathcal{G}_{\mu {\nu }%
}=$ Re$Q_{\mu {\nu }}$, where the geometric tensor $Q_{\mu {\nu }}$ can be
obtained by substituting Eq. (\ref{Bloch wave}) to Eq. (\ref{qgt}), and it
can be verified that this metric $\mathcal{G}$ is given by the following
diagonalized form \cite{note1}
\begin{equation}
dS^{2}=\mathcal{G}_{kk}dk^{2}+\mathcal{G}_{\varphi \varphi }d\varphi ^{2},
\end{equation}%
with%
\begin{eqnarray}
\mathcal{G}_{kk} &=&\left[ \frac{1}{2}\frac{\gamma +\gamma (h-2\delta
+\delta \text{cos}2k)\text{cos}k}{(h+\text{cos}k-\delta \text{cos}%
2k)^{2}+\gamma ^{2}\text{sin}^{2}k}\right] ^{2},  \notag \\
\mathcal{G}_{\varphi \varphi } &=&\frac{\gamma ^{2}\text{sin}^{2}k}{(h+\text{%
cos}k-\delta \text{cos}2k)^{2}+\gamma ^{2}\text{sin}^{2}k}.
\label{metrictensor}
\end{eqnarray}
The metric $\mathcal{G}$ is obviously independent on the parameter $\varphi $
because of its $U(1)$ gauge invariance on the twist operator.
\begin{figure}[t]
\begin{center}
\includegraphics[width=1.65in]{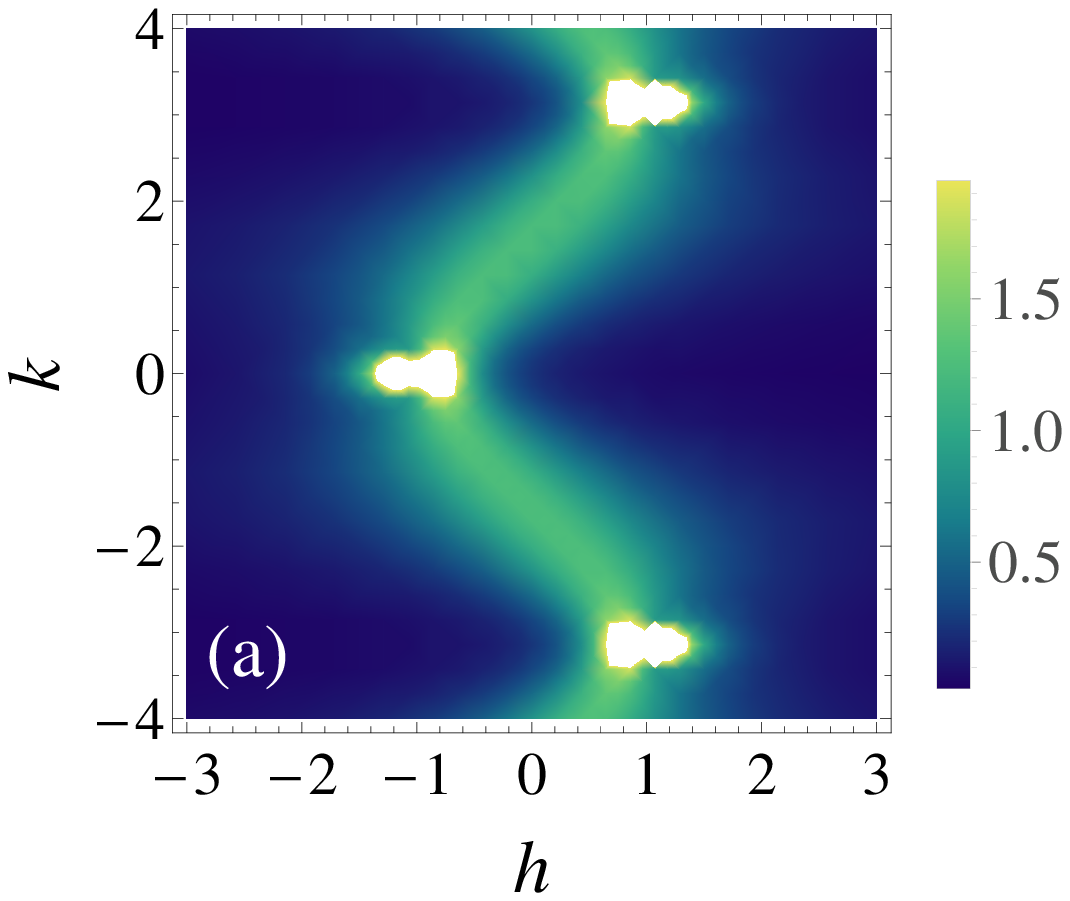} %
\includegraphics[width=1.65in]{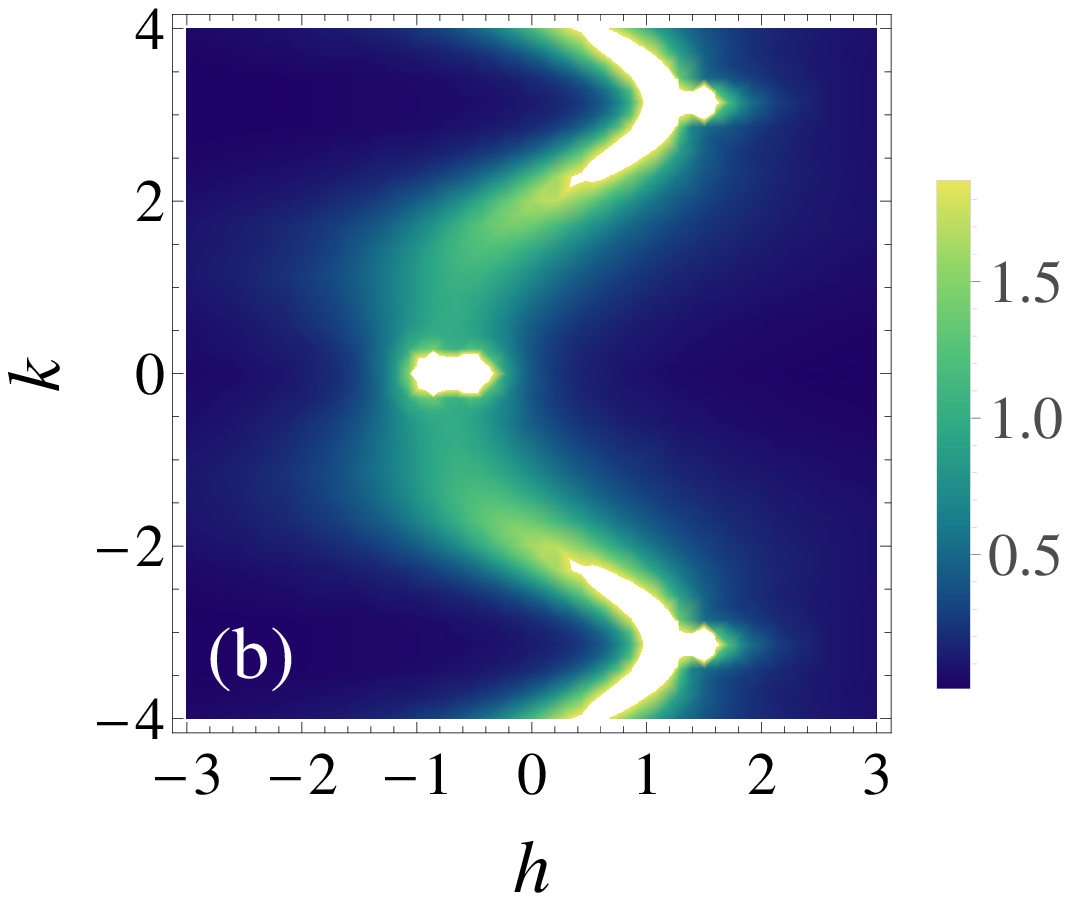}
\end{center}
\caption{(color online) The trace of the Riemannian metric $\text{Tr}\mathcal{G}$
as a function of the external field $h$ and the crystal momentum $k$ with
the fixed Hamiltonian parameters: (a) the three-site spins exchange
interactions $\protect\delta =0$ and the anisotropy parameter $\protect%
\gamma =1$; (b) $\protect\delta =0.3$ and $\protect\gamma =0.9$.}
\label{trace}
\end{figure}

\begin{figure}[tbh]
\includegraphics[width=3.3in]{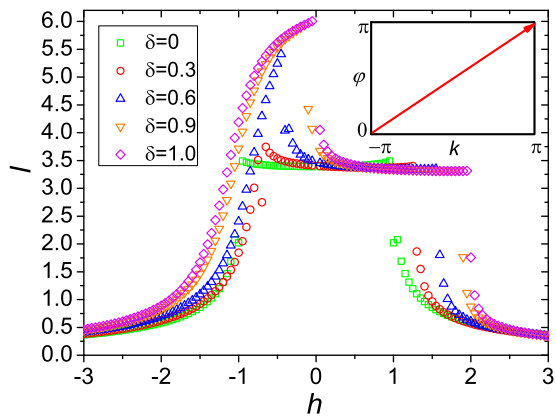}
\caption{(color online) The cyclic quantum distance $l$ of the Bloch band as
a function of $h$, with the fixed anisotropy parameter $\protect\gamma =1/3$
and different three-site spins coupled coefficients $\protect\delta $, where
the integral path is along the diagonal line in the extended Brillouin zone
(inset).}
\end{figure}

\begin{figure}[tbh]
\includegraphics[width=3.3in]{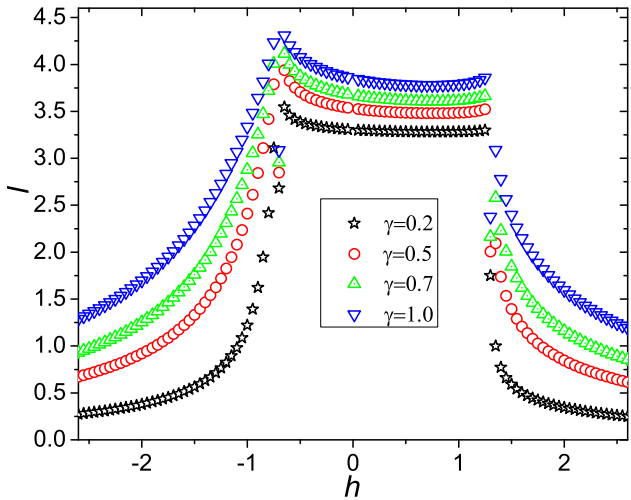}
\caption{(color online) The cyclic quantum distance $l$ of the Bloch band as
a function of $h$, with the fixed three-site spins coupled coefficient $%
\protect\delta=0.3$ and different anisotropy parameters $\protect\gamma $.}
\end{figure}

In Fig. 1, we show that the trace of the Riemannian metric as a function of
the external field $h$ and the crystal momentum $k$ with different
three-site spins coupled parameters and the anisotropy parameters. As we
expect, the singularity regions of the metric will appear when the external
field $h$ is close to the quantum critical points. We define a cyclic quantum
distance on the Bloch band from $\left( 0,-\pi \right) $ to $\left( \pi ,\pi
\right) $ in the extended first Brillouin zone (the inset in Fig.2), where
the parameter path of the integral loop $C$ is $\varphi =k/2+\pi /2$, $%
\left( k\in 1\text{Bz}\right) $, which is just the diagonal line in the extended
Brillouin zone.

The cyclic quantum distance $l$ of the Bloch band is given by
\begin{equation}
l=\oint_{C}\sqrt{\mathcal{G}_{kk}dk^{2}+\mathcal{G}_{\varphi \varphi
}d\varphi ^{2}}=\int_{-\pi }^{\pi }\sqrt{\mathcal{G}_{kk}+\frac{1}{4}%
\mathcal{G}_{\varphi \varphi }}d{k}.  \label{eq8}
\end{equation}%
where the Riemannian metric $\mathcal{G}_{kk}$ and $\mathcal{G}_{\varphi
\varphi }$ are given by Eq. (\ref{metrictensor}). As shown in Fig. 2, we
calculate the cyclic quantum distance $l$ as a function of $h$, with the
fixed anisotropy parameter $\gamma =1/3$ and different three-site spins
coupled coefficients $\delta $. The singularity points on the cyclic quantum
distance are just corresponding to the quantum transition points $|\delta-1
| $ and $|\delta+1|$.

In Fig. 3, we plot the the cyclic quantum distance $l$ with the fixed
three-site spins coupled coefficients $\delta=0.3$ and different anisotropy
parameters $\gamma $. It can be seen that the value of the anisotropy
parameter $\gamma $ does not affect the critical point but makes the cyclic
quantum distance $l$ approach zero more quickly in the paramagnetic phase.

\subsection{Cyclic quantum distance of the ground state}

It is worth noting that the metric component $\mathcal{G}_{\varphi \varphi }$
on the Bloch band is closely related to the ground-state quantum distance in
the parameter $\varphi $ space. In fact, the ground state $\left\vert {%
e(\varphi )}\right\rangle $ is $\pi $ periodic in the parameter $\varphi $.
In the condition of the large sites limit $N\rightarrow \infty $, a cyclic
ground-state distance $l_{e}$ can be defined along the $\varphi $ -ring as
\begin{eqnarray}
l_{e} &=&\int_{0}^{\pi }\sqrt{\langle {{{\partial _{\varphi }e(\varphi )}}}%
|\left[ \boldsymbol{1-}\mathcal{P}_{e}\right] \left\vert {\partial }_{{%
\varphi }}{e(\varphi )}\right\rangle }d{\varphi }  \notag \\
&=&\frac{1}{2\pi }\iint \sqrt{\mathcal{G}_{\varphi \varphi }}d{k}d{\varphi ,}
\label{gsmetric}
\end{eqnarray}%
where $\mathcal{P}_{e}=\left\vert {e}\left( \varphi \right) \right\rangle
\left\langle {e}\left( \varphi \right) \right\vert $ denotes the
ground-state projection operator and the Eqs. (\ref{Bloch wave}) and (\ref%
{gs}) have been used in the intermediate steps. Note that the result in Eq. (%
\ref{gsmetric}) is general, which only relates to the metric $\mathcal{G}%
_{\varphi \varphi }$ on the Bloch band and the concrete expression of the
ground state is not required.
\begin{figure}[tbh]
\begin{center}
\includegraphics[width=3.3in]{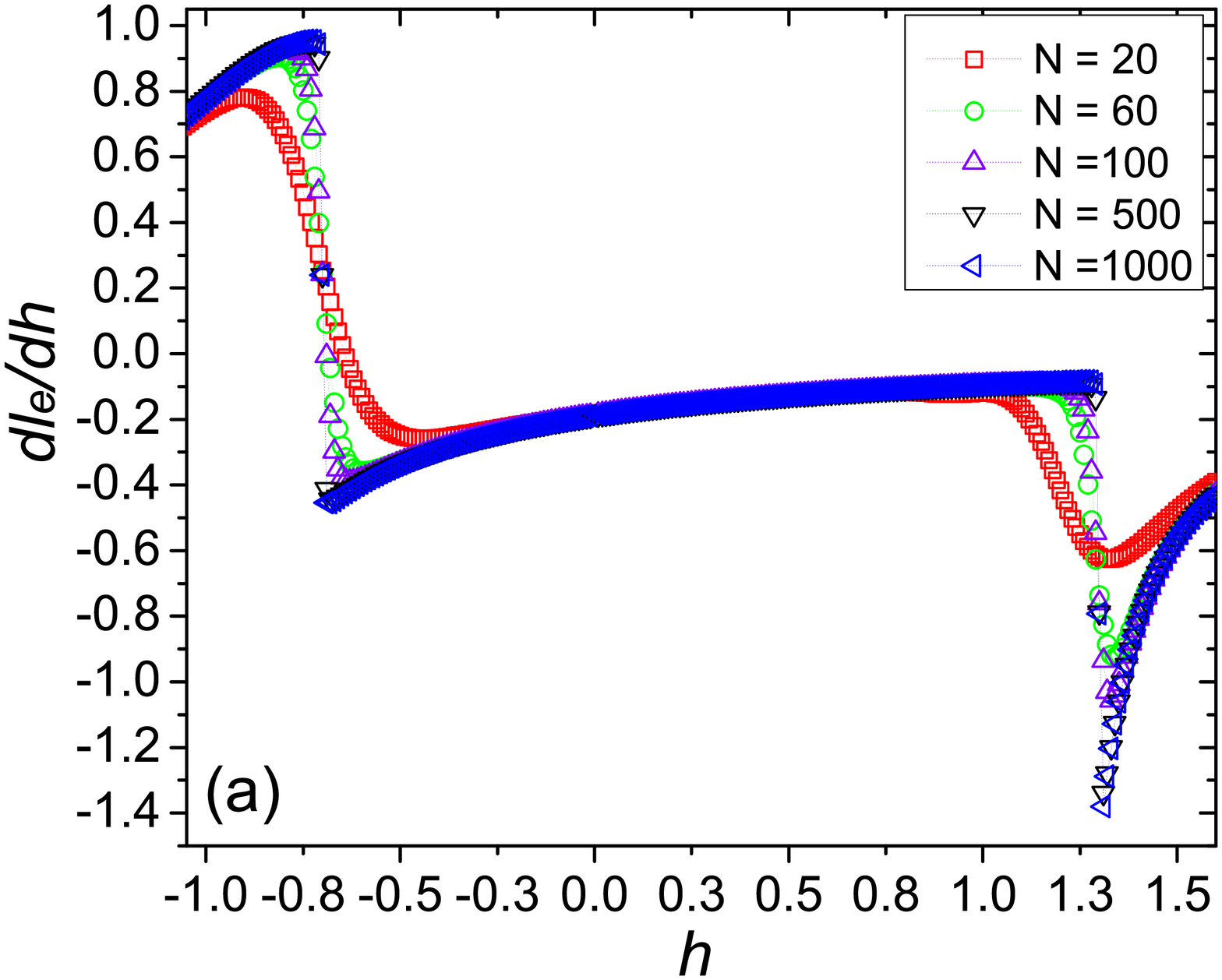} %
\includegraphics[width=3.3in]{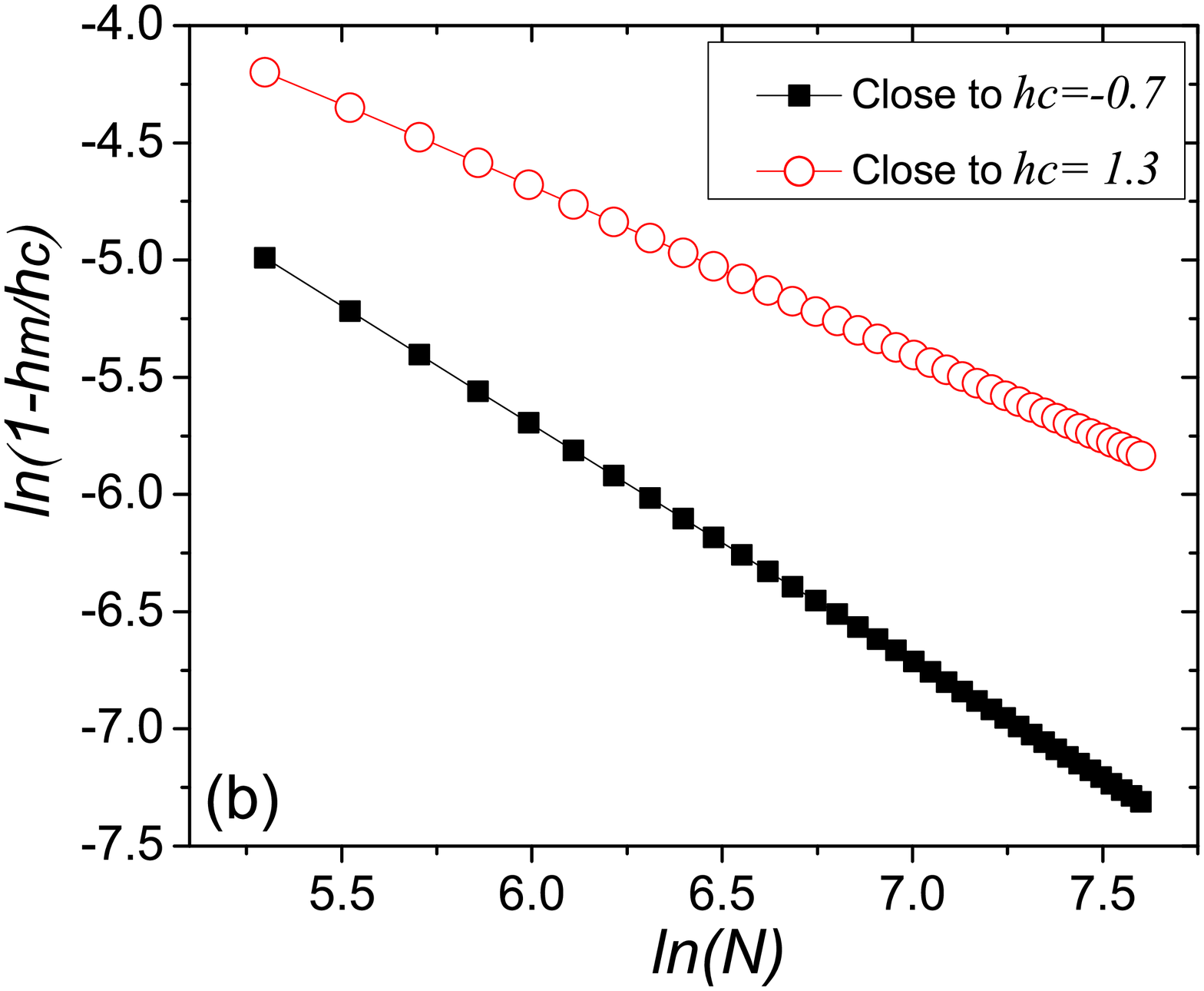}
\end{center}
\caption{(color online) (a) The derivative of the cyclic ground-state
distance $dl_{e}/dh$ with different lattice sizes $N$, where the
Hamiltonian parameters $\protect\gamma =0.7$ and $\protect\delta =0.3$; (b)
with the increasing of the lattice sizes, the positions of the maximum
points of $dl_{e}/dh$ tend as $N^{-1.0074}$ and $N^{-0.7122}$ to the
critical points $h_{c}=0.7$ and $h_{c}=1.3$, respectively.}
\end{figure}

In Fig. 4 (a), we show the derivative of the cyclic ground-state distance
with respect to the external field $h$ under different lattice sizes $N$,
where the Hamiltonian parameters $\gamma =0.7$ and $\delta =0.3$. As shown
in Fig. 4 (b), we can see that the positions of the maximum points of the
derivative $dl_{e}/dh$, with the increasing of the lattice sizes, tend as $%
N^{-1.0074}$ and $N^{-0.7122}$ to the critical points $h_{c}=0.7$ and $%
h_{c}=1.3$, respectively.

\section{The Euler Characteristic Number of the Bloch Band}

What is more interesting is that the Euler characteristic number of the
Bloch band can be derived from the Gauss-Bonnet theorem on the Bloch states
manifold established by the Riemannian metric $\mathcal{G}_{\mu {\nu }%
}^{\left( n\right) }$. The Euler characteristic number $\chi $\ of all
occupied bands can be generalized written by (see Ref. \cite{Ma2013})
\begin{equation}
\chi =\frac{1}{4\pi }\sum_{n}\iint_{1\text{Bz}}\mathcal{R}^{\left( n\right) }%
\sqrt{\det \mathcal{G}_{\mu {\nu }}^{\left( n\right) }}d{k}^{\mu }d{k}^{{\nu
}},  \label{eu}
\end{equation}%
where the $\mathcal{R}^{\left( n\right) }$ is the Ricci scalar curvature
associate to the Bloch state $\left\vert {u}_{n}{\left( k\right) }%
\right\rangle $ of the $n$-th Bloch band. The Ricci scalar curvature $%
\mathcal{R}$ can be calculated by using the standard steps: $\mathcal{R}=%
\mathcal{G}^{ab}R_{acb\cdot }^{\text{ }\cdot \cdot \cdot c}$, where the
Riemannian curvature tensor
\begin{equation}
R_{abc\cdot }^{\text{ }\cdot \cdot \cdot d}=\partial _{b}\Gamma
_{ac}^{d}-\partial _{a}\Gamma _{bc}^{d}+\Gamma _{ac}^{e}\Gamma
_{be}^{d}-\Gamma _{bc}^{e}\Gamma _{ae}^{d},
\end{equation}%
and the Levi-Civit\`{a} connection $\Gamma _{bc}^{a}$ can be calculated by
\begin{equation}
\Gamma _{bc}^{a}=\frac{1}{2}\mathcal{G}^{ad}\left( {\partial }_{b}\mathcal{G}%
_{dc}+{\partial }_{c}\mathcal{G}_{bd}-{\partial }_{d}\mathcal{G}_{cb}\right)
.  \label{conn}
\end{equation}%
The Riemannian metric $\mathcal{G}$ of the Bloch band is given by Eq. (\ref%
{metrictensor}), and its contravariant component can be easily obtained as $%
\mathcal{G}^{kk}=1/\mathcal{G}_{kk}$, and $\mathcal{G}^{\varphi \varphi }=1/%
\mathcal{G}_{\varphi \varphi }$. By using Eqs. (\ref{metrictensor}) and (\ref%
{conn}), we can obtain all of the non-zero connections as 
\begin{eqnarray}
\Gamma _{k\varphi }^{\varphi } &=&\Gamma _{\varphi k}^{\varphi }  \notag \\
&=&\frac{\left( B-\gamma ^{2}\text{cos}k\right) \text{sin}k-2\delta B\text{%
sin}2k}{B^{2}+\gamma ^{2}\text{sin}^{2}k}  \notag \\
&&+\cot k,  \notag \\
\Gamma _{kk}^{k} &=&2\frac{\left( B-\gamma ^{2}\text{cos}k\right) \text{sin}%
k-2\delta B\text{sin}2k}{B^{2}+\gamma ^{2}\text{sin}^{2}k}  \notag \\
&&-\frac{(h+3\delta \text{cos}2k)\text{sin}k}{1+A\text{cos}k},  \notag \\
\Gamma _{\varphi \varphi }^{k} &=&-\frac{4B\text{sin}k}{1+A\text{cos}k},
\end{eqnarray}%
%
%
%
%
%
%
%
%
%
%
%
%
with
\begin{eqnarray}
A &=&h+\delta \text{cos}2k-2\delta ,  \notag \\
B &=&h-\delta \text{cos}2k+\text{cos}k.
\end{eqnarray}

\begin{figure}[tbh]
\includegraphics[width=3.3in]{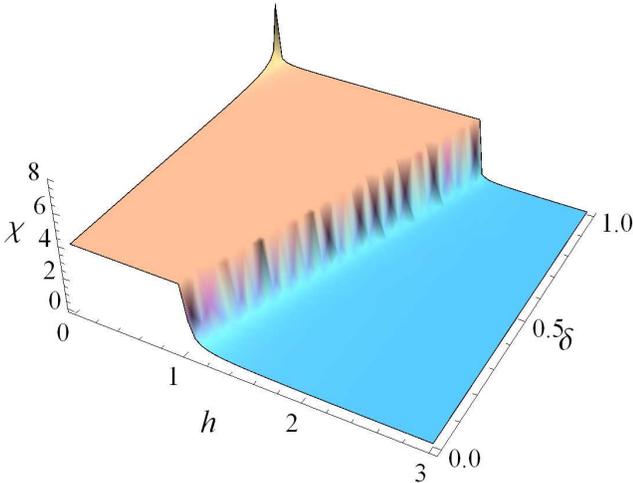} 
\caption{(color online) The Euler number $\protect\chi$ of the Bloch band as
a function of the external field $h$ and three-site spins coupled
coefficients $\protect\delta$. The ferromagnetic phase in this model can be
marked by a nontrivial Euler number $\protect\chi =4$, and the Euler number $%
\protect\chi \rightarrow 0$ quickly with the increasing of the external
field $h$ in the paramagnetic phase.}
\end{figure}

\begin{figure}[tbh]
\includegraphics[width=3.3in]{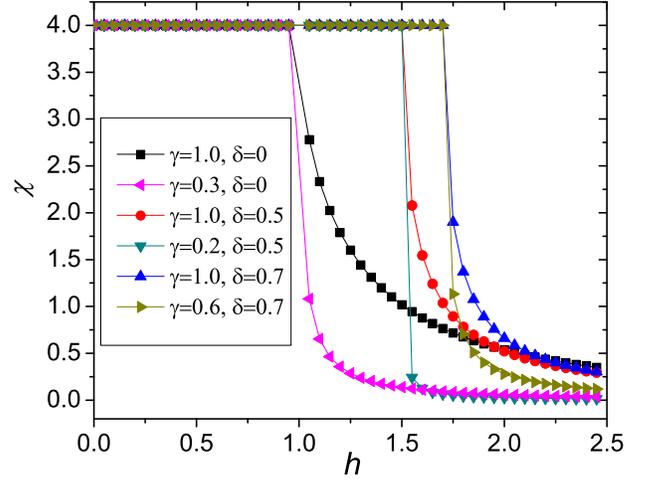} 
\caption{(color online) The Euler number $\protect\chi$ with several groups
of the anisotropy parameter $\protect\gamma$ and three-site spins coupled
coefficients $\protect\delta$.}
\end{figure}

The Euler characteristic number $\chi $ is a topological invariant and
equals to $2\left( 1-g\right) $ with genus $g$ for a closed smooth manifold.
Note that the Bloch band of the model forms a 2D closed Riemannian manifold
in the first Brillouin zone, and then the Euler characteristic number can be
calculated conveniently by the Gauss-Bonnet theorem $\chi =\frac{1}{2\pi }%
\int_{1\text{Bz}}\mathcal{K}dA$, where $\mathcal{K=}R_{k\varphi \varphi k}^{%
\text{ \ \ }}/\det \mathcal{G}_{k\varphi }$ is the Gauss curvature, which
just equals to the half of the Ricci scalar curvature $\mathcal{R}$, and the
covariant Riemannian curvature tensor $R_{abcd}^{\text{ \ \ }}:=R_{abc\cdot
}^{\text{ }\cdot \cdot \cdot e}\mathcal{G}_{ed}$ have only one substantial
component $R_{k\varphi k\varphi }^{\text{ \ \ }}$, and $dA=\sqrt{\det
\mathcal{G}_{k\varphi }}d{k}d\varphi $ denotes the area measure.

The direct calculation of the $R_{k\varphi k\varphi }^{\text{ \ \ }}$ and $%
\sqrt{\det \mathcal{G}_{k\varphi }}$ are tedious, however, it can be
verified that there exists a general relation in a generalized two-band
Hamiltonian on a 2D manifold as $R_{k\varphi k\varphi }^{\text{ \ \ }}=4\det
\mathcal{G}_{k\varphi }$ and $\det \mathcal{G}_{k\varphi }=\left( \frac{%
\boldsymbol{\hat{d}}\cdot {\partial _{k}}\boldsymbol{\hat{d}}\times {%
\partial _{\varphi }}\boldsymbol{\hat{d}}}{4}\right) ^{2}$. That is to say
that the Bloch band manifold is a curved surface with a constant Gauss
curvature $\mathcal{K}=4$. Finally, we can derive the Euler number of the
Bloch band as
\begin{eqnarray}
\chi &=&\frac{1}{2\pi }\int_{1\text{Bz}}\mathcal{K}dA  \notag \\
&=&\frac{1}{2\pi }\iint_{1\text{Bz}}\left\vert \boldsymbol{\hat{d}}\cdot {%
\partial _{k}}\boldsymbol{\hat{d}}\times {\partial _{\varphi }}\boldsymbol{%
\hat{d}}\right\vert d{k}d\varphi ,  \label{eu2}
\end{eqnarray}%
where
\begin{equation}
\boldsymbol{\hat{d}}\cdot {\partial _{k}}\boldsymbol{\hat{d}}\times {%
\partial _{\varphi }}\boldsymbol{\hat{d}}=\frac{2\gamma ^{2}\text{sin}%
k+\gamma ^{2}(h-2\delta +\delta \text{cos}2k)\text{sin2}k}{\left[ (h+\text{%
cos}k-\delta \text{cos}2k)^{2}+\gamma ^{2}\text{sin}^{2}k\right] ^{3/2}}.
\end{equation}

As shown in Fig. 5, we plot the Euler number $\chi $ of the Bloch band as a
function of the external field $h$ and three-site spins coupled coefficients
$\delta $. In the ferromagnetic phase, the Bloch band is characterized by a
nontrivial Euler number $\chi =4$, whose topology is equivalent to two
unconnected spheres $S^{2}$; in the paramagnetic phase, the Euler number of
the Bloch band $\chi \rightarrow 0$ quickly with the increasing of the
external field $h$, whose topology is equivalent to a torus $T^{2}$. The
effects of the anisotropy parameter $\gamma $ on the Euler number are shown
in the Fig. 6. It can be seen that the Euler number is independent of $%
\gamma $ in the region of the ferromagnetic phase, but declines to $0$ more
quickly with the decreasing of $\gamma $ in the region of the paramagnetic
phase.

Note that the Berry curvature of the Bloch band can be written as $\mathcal{F%
}_{k}{_{\varphi }}=\frac{1}{2}\boldsymbol{\hat{d}}\cdot {\partial _{k}}%
\boldsymbol{\hat{d}}\times {\partial _{\varphi }}\boldsymbol{\hat{d}}$, so
we can get a first Chern number index for the Bloch band as%
\begin{equation}
C_{1}=\frac{1}{4\pi }\iint_{1\text{Bz}}\boldsymbol{\hat{d}}\cdot {\partial
_{k}}\boldsymbol{\hat{d}}\times {\partial _{\varphi }}\boldsymbol{\hat{d}}%
\text{ }d{k}d\varphi .  \label{Chern number}
\end{equation}%
However, the Bloch Hamiltonian for this model $\mathcal{H}(k,\varphi
)=\sum_{\alpha =1}^{3}d_{\alpha }\left( k,\varphi \right) \sigma ^{\alpha }$
is time reversal invariant, i.e. $\mathcal{H}^{\ast }(-k,-\varphi )=\mathcal{%
H}(k,\varphi )$, so the Berry curvature $\mathcal{F}_{k}{_{\varphi }}$ is
odd with the crystal momentum $k$ (note $\mathcal{F}_{k}{_{\varphi }}$ is
not dependent on $\varphi $), and the first Chern number $C_{1}\equiv 0$. In
this case, the first Chern number can not serve as a sufficient index for
the topology of the Bloch band in the time reversal invariant systems.

\begin{figure}[thb]
\begin{center}
\includegraphics[width=0.23\textwidth,height=0.15\textheight]{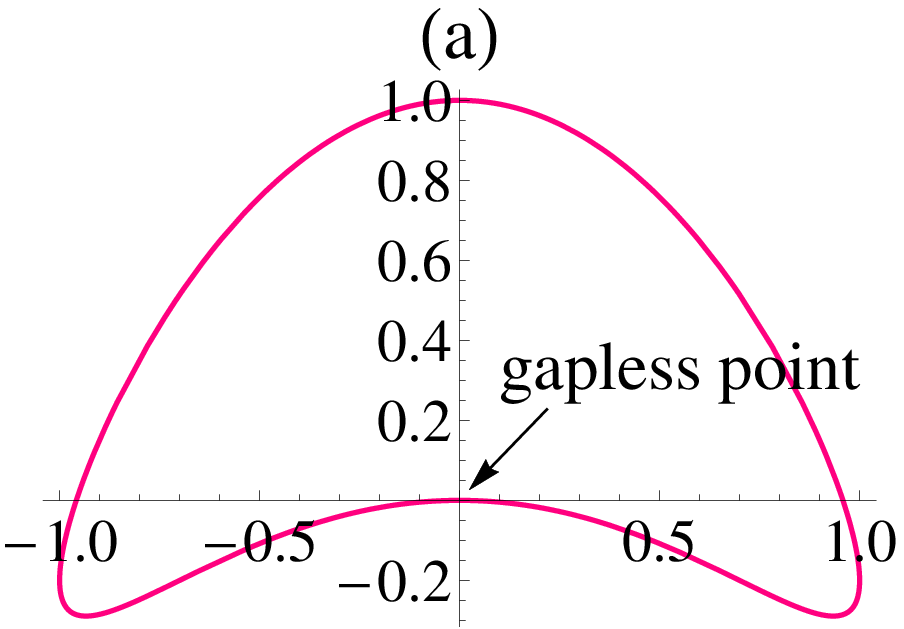} %
\includegraphics[width=0.23\textwidth,height=0.15\textheight]{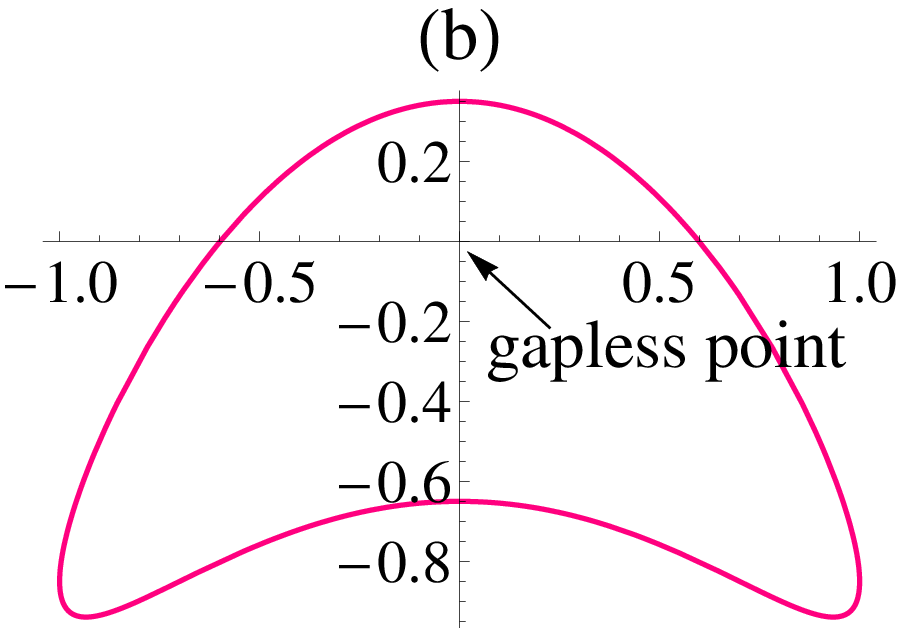} %
\includegraphics[width=0.23\textwidth,height=0.15\textheight]{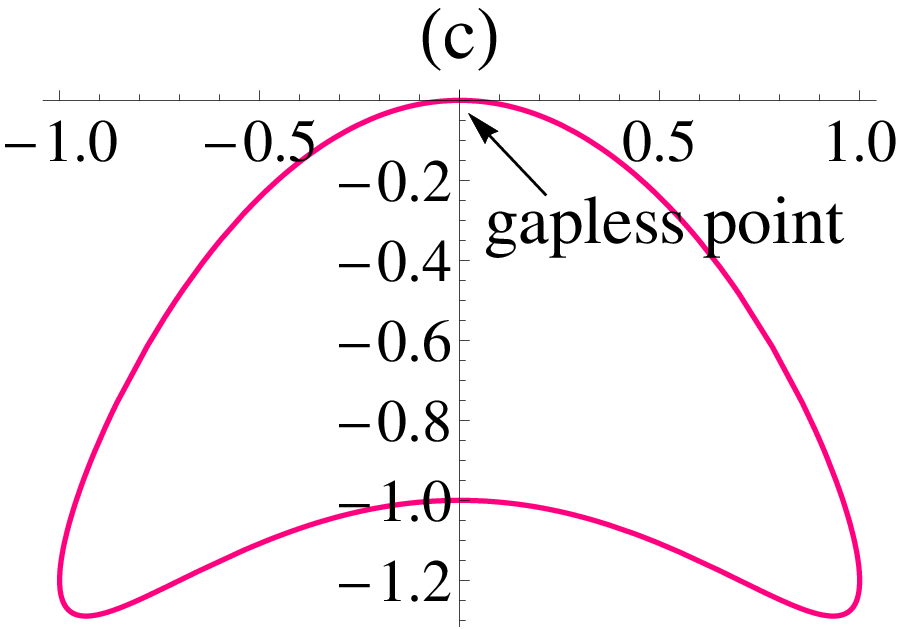} %
\includegraphics[width=0.23\textwidth,height=0.15\textheight]{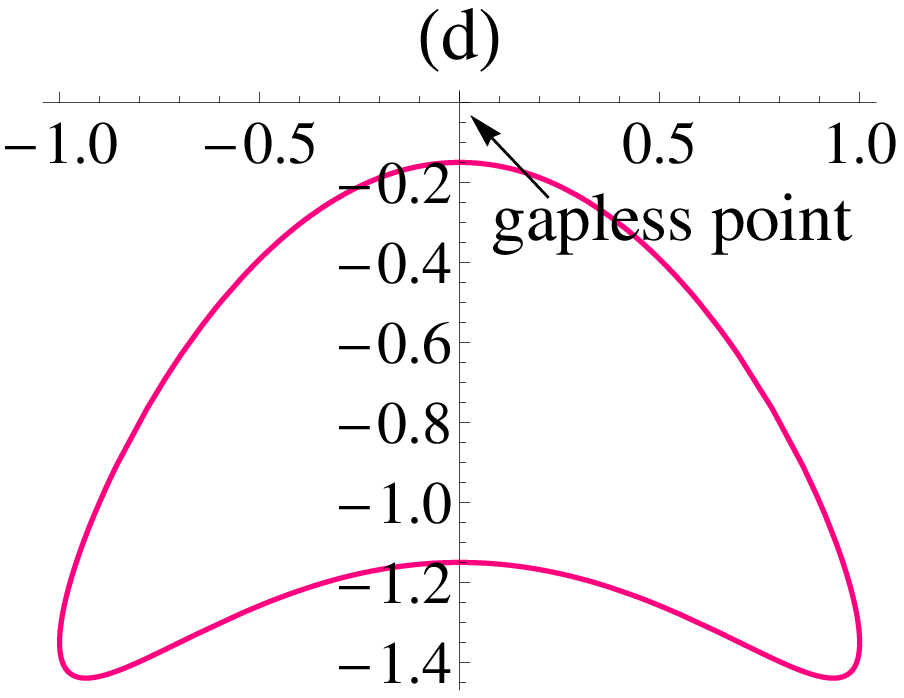}
\end{center}
\caption{(color online) The trajectories of the 2D vector $%
\boldsymbol{d}(k)$ with the Hamiltonian parameters $\protect\delta =0.7$, $%
\protect\gamma =1$ and (a) $h=-0.3$; (b) $h=1$; (c) $h=1.7$; (d) $h=2$.}
\label{2dcur}
\end{figure}

For an intuitive picture, the original 1D fermionic Hamiltonian without the
twist operation is given by
\begin{equation}
H_{1D}=\sum_{k\in \text{Bz}}\left( c_{k}^{\dagger }\text{, }c_{-k}\right)
\mathcal{H}_{\text{1D}}(k)\left(
\begin{array}{c}
c_{k} \\
c_{-k}^{\dagger }%
\end{array}%
\right) ,  \label{h1d}
\end{equation}%
where the Bloch Hamiltonian $\mathcal{H}_{\text{1D}}(k)=\sum_{\alpha
=1}^{3}d_{\alpha }\left( k\right) \sigma ^{\alpha }$, with $d_{1}\left(
k\right) =0$, $d_{2}\left( k\right) =\frac{1}{2}\gamma \sin k$, $d_{3}\left(
k\right) =\frac{1}{2}\left( -h+\delta \cos 2k-\cos k\right) $, can be
verified to be time-reversal invariant because $\mathcal{H}_{\text{1D}}^{\ast
}(-k)=\mathcal{H}_{\text{1D}}(k)$. Meanwhile, $\mathcal{H}_{\text{1D}}(k)$
has a particle-hole symmetry $\left( \sigma ^{x}K\right) \mathcal{H}_{\text{%
1D}}(k)\left( \sigma ^{x}K\right) ^{-1}=-\mathcal{H}_{\text{1D}}(-k)$
because $d_{3}\left( k\right) ^{\ast }=$ $d_{3}\left( -k\right) $, $%
d_{1}\left( k\right) ^{\ast }=-d_{1}\left( -k\right) $ and $d_{2}\left(
k\right) ^{\ast }=-d_{2}\left( -k\right) $. As a result of the time-reversal
and particle-hole symmetry, the Hamiltonian $\mathcal{H}_{\text{1D}}(k)$ has
also a chiral symmetry. Therefore, the 1D Hamiltonian $H_{1D}$ is in the BDI
class. It has been shown that BDI class Hamiltonians in one dimension are
classified by an integer $\mathbb{Z}$ topological invariant \cite{topotable}. The $\mathbb{Z}$
number can be expressed as the winding number of the 2D vector $\boldsymbol{d%
}(k)=\left( d_{2}\left( k\right) ,d_{3}\left( k\right) \right) $ around the
gapless point $\left\vert \boldsymbol{d}(k)\right\vert =0$ when $k$ runs
across the first Brillouin zone. The winding number can be written as \cite%
{BDI}
\begin{equation}
N_{\text{BDI}}=\frac{1}{2\pi }\int_{-\pi }^{\pi }d\Phi \left( k\right) ,
\end{equation}%
where $\Phi \left( k\right) =\arctan \left[ d_{3}\left( k\right)
/d_{2}\left( k\right) \right] $ denotes the angle of the vector $\boldsymbol{%
d}(k)$. As shown in Fig. \ref{2dcur}, we plot the vector $\boldsymbol{d}%
(k) $ with the Hamiltonian parameters $\delta =0.7$, $\gamma =1$ and
different $h $. It can be seen clearly that the ferromagnetic phase and
paramagnetic phase are topologically nonequivalent depending on whether or not
the gapless point is enclosed within the the curve of $\boldsymbol{d}(k)$,
and the quantum phase transitions occur at $h=\delta \pm 1$.

\begin{figure}[thb]
\begin{center}
\includegraphics[width=0.23\textwidth,height=0.15\textheight]{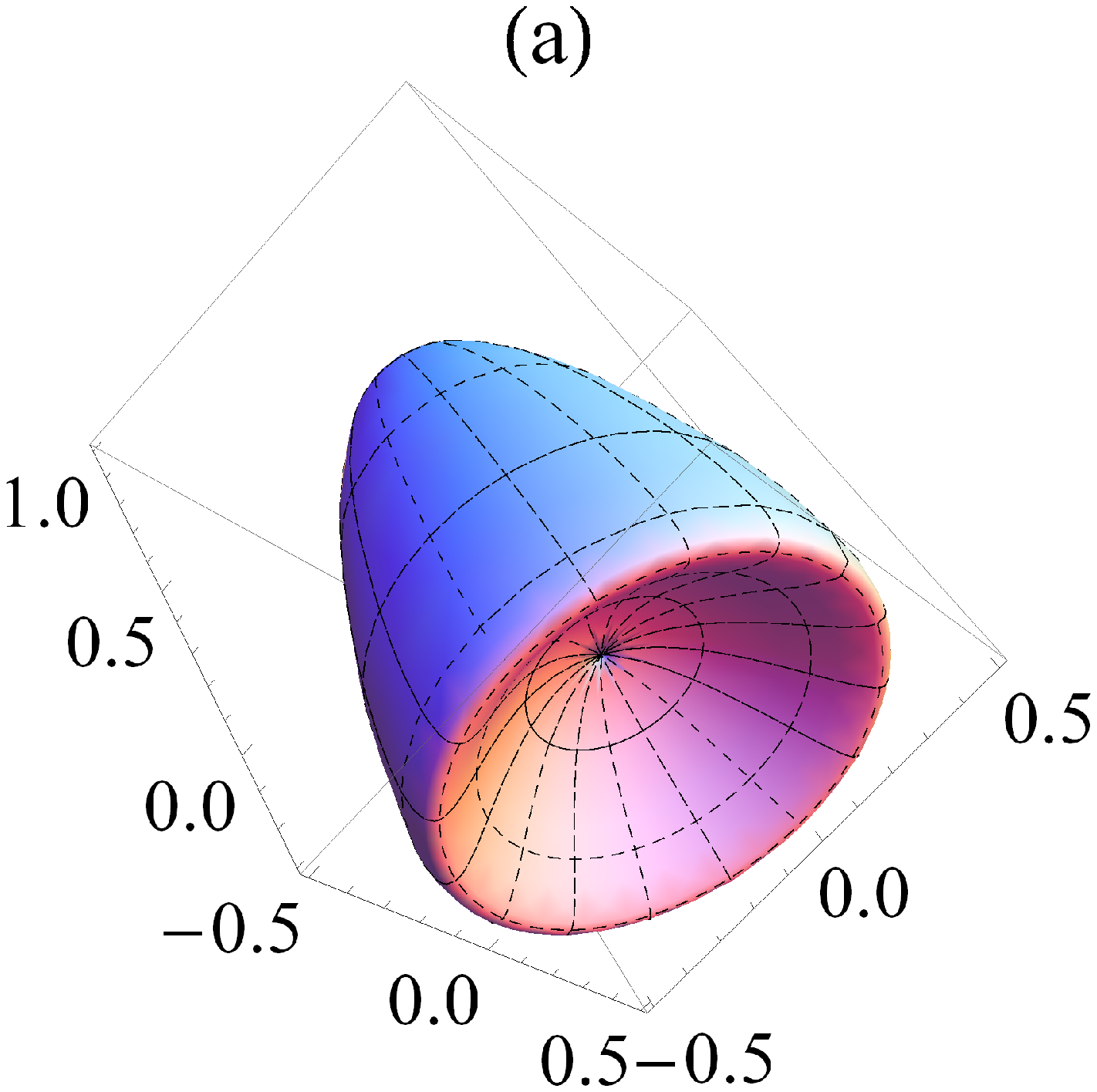} %
\includegraphics[width=0.23\textwidth,height=0.15\textheight]{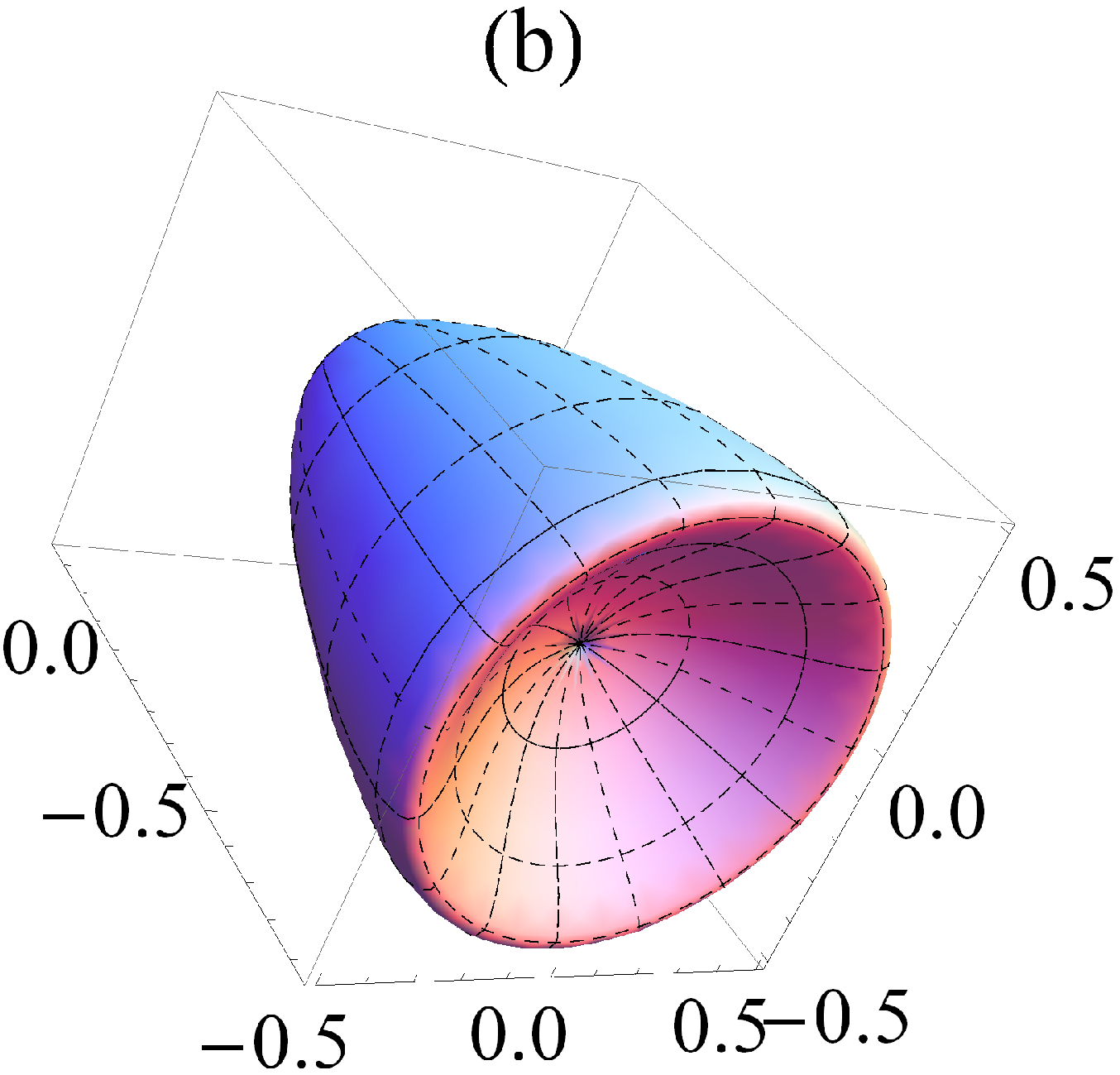} %
\includegraphics[width=0.23\textwidth,height=0.15\textheight]{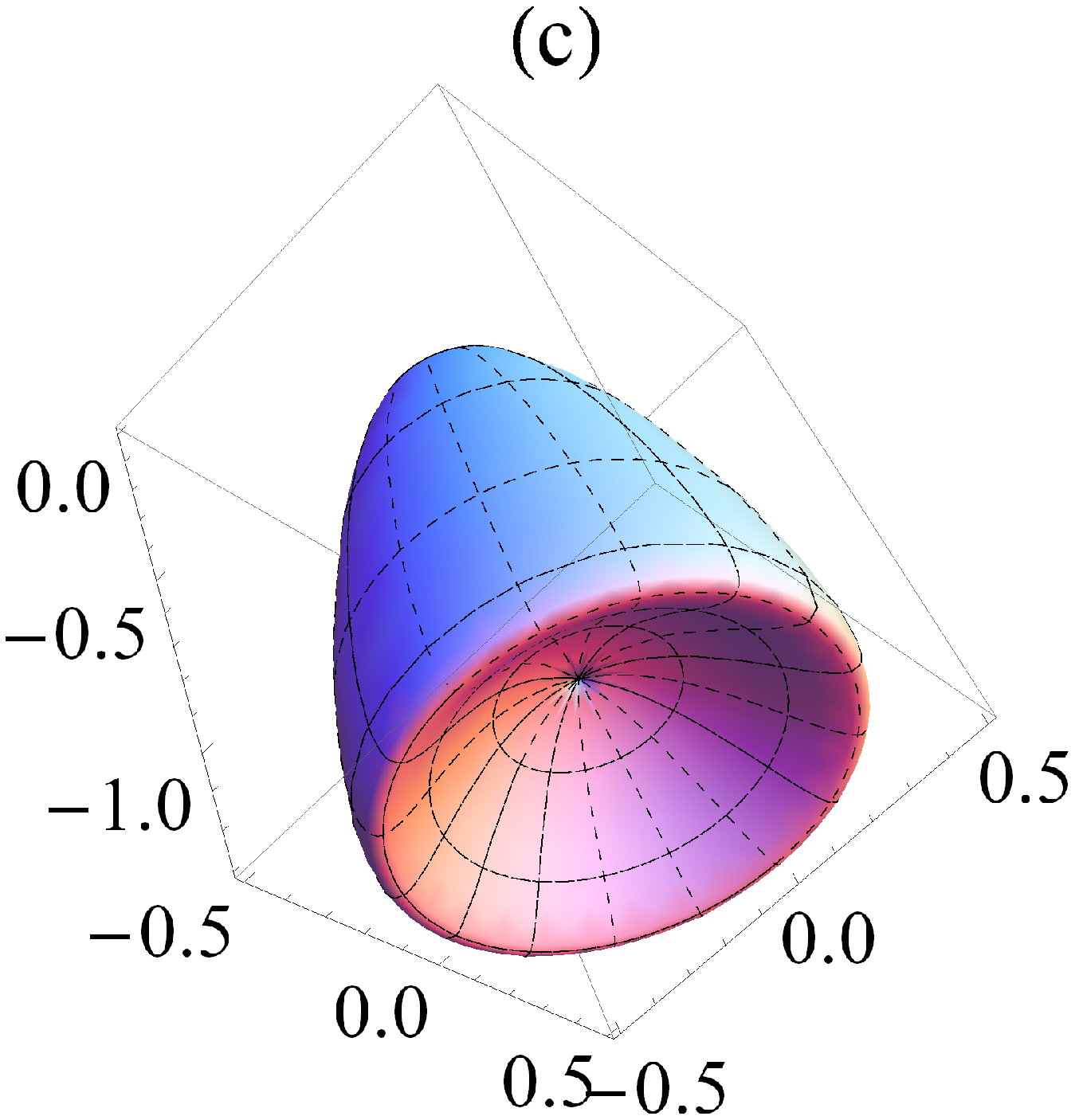} %
\includegraphics[width=0.23\textwidth,height=0.15\textheight]{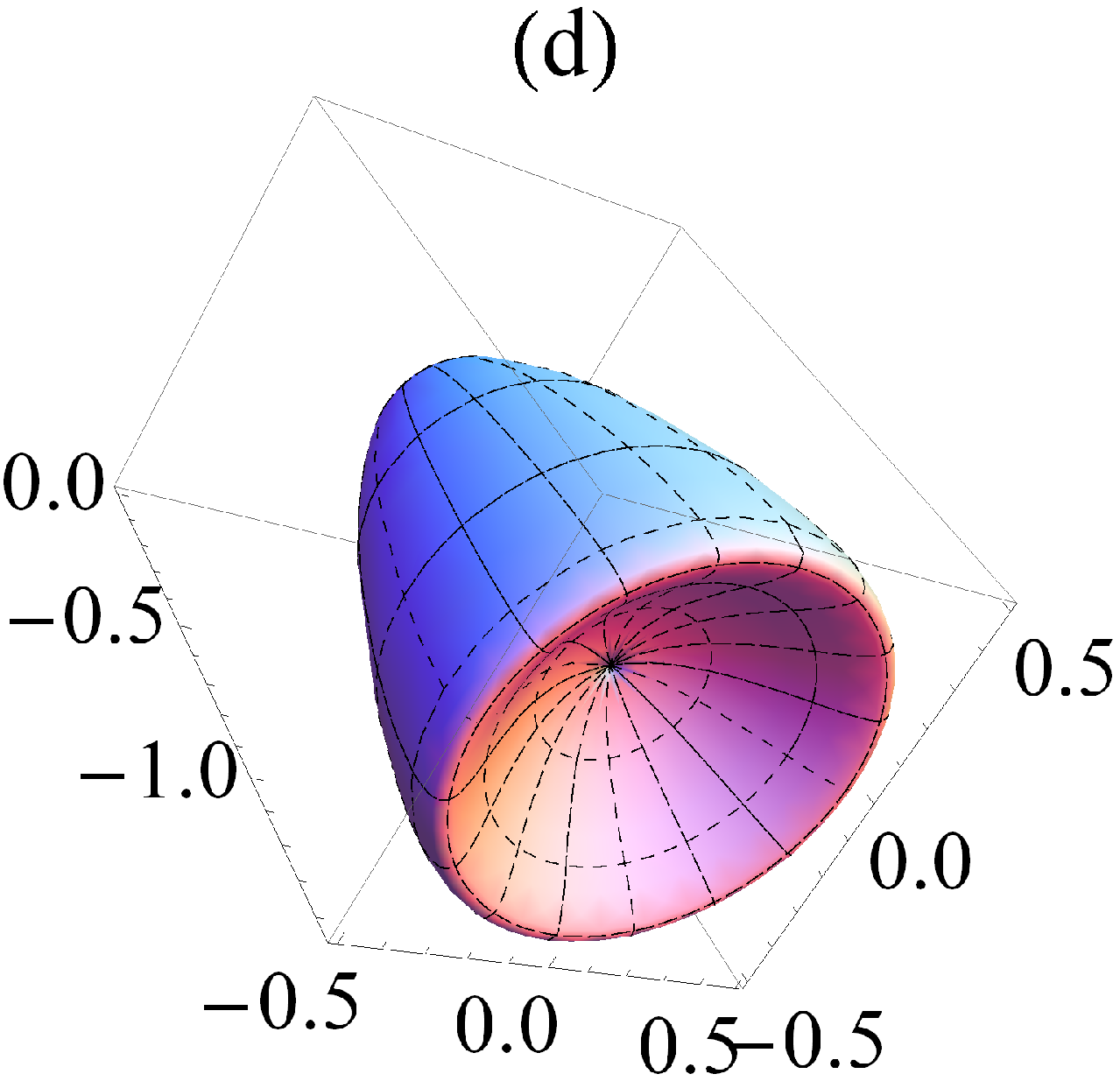}
\end{center}
\caption{(color online) The trajectories of the 3D vector $%
\boldsymbol{d}(k,\protect\varphi )$ with the Hamiltonian parameters $\protect%
\delta =0.7$, $\protect\gamma =1$ and (a) $h=-0.3$; (b) $h=1$; (c) $h=1.7$; (d) $h=2$.}
\label{3dcur}
\end{figure}

In the Euler number approach, the 1D Hamiltonian has been extended to two dimensions by
subjecting the system to a gauge transformation (see Eq. (\ref{ferham})),
and meanwhile its energy spectrum remains unchanged. As a consequence, the
2D Bloch Hamiltonian $\mathcal{H}(k,\varphi )$ only remains
time-reversal invariant, and then belongs to the symmetry class AI without
a strong\ topological invariant in two dimensions.

However, it can be verified that there exists an intuitive topological
connection between $\mathcal{H}(k,\varphi )$ and $\mathcal{H}_{\text{1D}}(k)$%
. As shown in Fig. \ref{3dcur}, the trajectories of the vector $%
\boldsymbol{d}(k,\varphi )$ are corresponding to the rotation of the vector $%
\boldsymbol{d}(k)$ around the ``$d_{3}$''-axis, and meanwhile, keeping the
gapless point unchanged. As the same as the 1D case, the ferromagnetic phase
and paramagnetic phase are topologically nonequivalent which can be
characterized by whether or not the gapless point is enclosed by the closed
surface of $\boldsymbol{d}(k,\varphi )$. Note that here the first Chern
number can not provide an effective distinction because the Berry curvature
is odd with $k$ in the time-reversal invariant Hamiltonian $\mathcal{H}%
(k,\varphi )$. However, we show that here the Euler number of the band can
servers as an effective topological number as the replacement of the Chern
number, because the Euler number can be expressed as the integral of the
absolute value of the Berry curvature in the first Brillouin zone.

It also needs to be pointed out that the Euler number in the ferromagnetic
phase is characterized by the even number $\chi=4$ instead of $\chi=2$.
\begin{figure}[tbh]
\begin{center}
\includegraphics[width=0.23\textwidth,height=0.15\textheight]{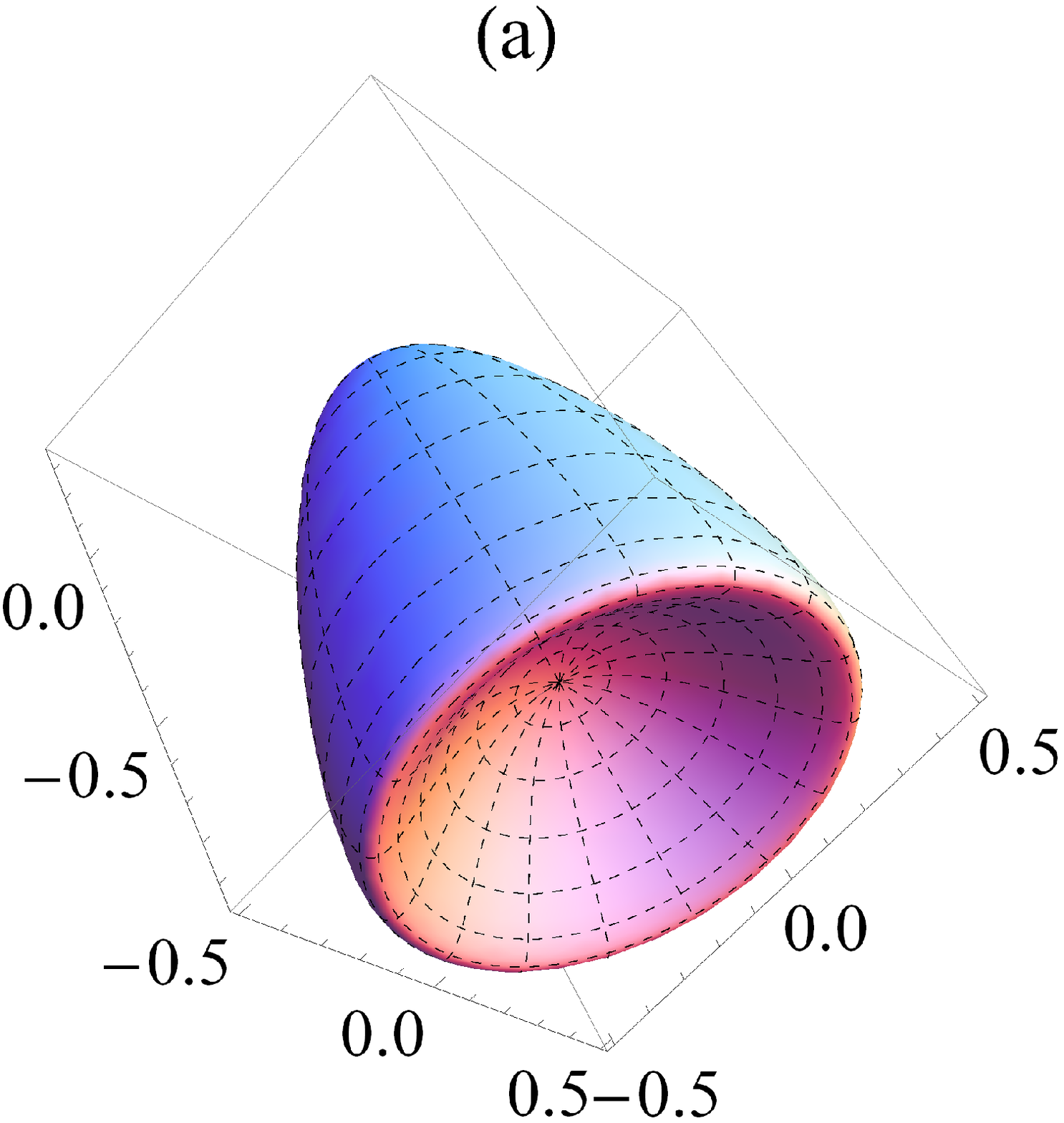} %
\includegraphics[width=0.23\textwidth,height=0.15\textheight]{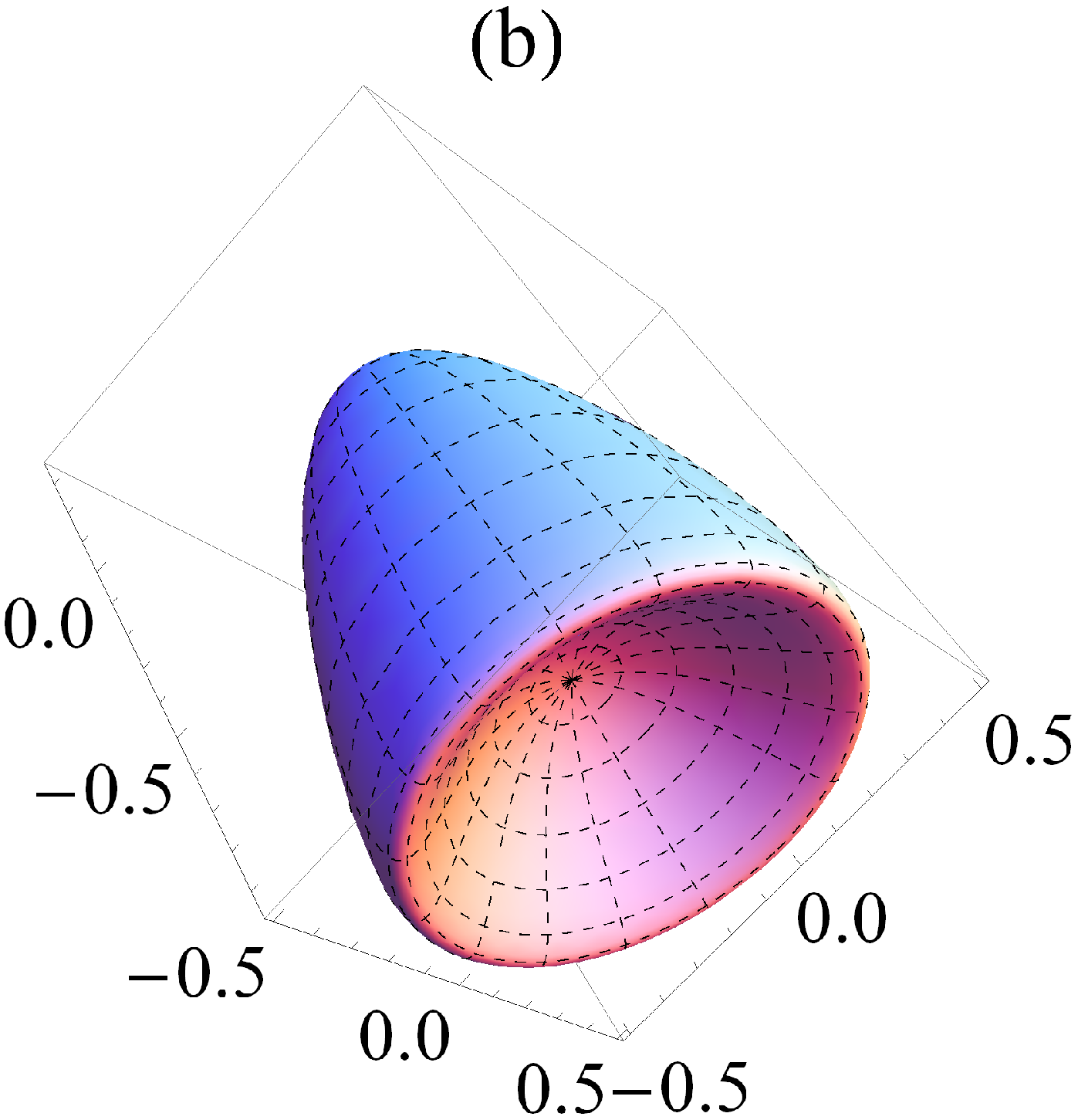}
\end{center}
\caption{(color online) The trajectories of the 3D vector $%
\boldsymbol{d}(k,\protect\varphi )$ with the Hamiltonian parameters $\protect\delta =0.7$, $\protect\gamma =1$ and $h=1$ can be split into two equal disjoint
topological spheres (a) $\boldsymbol{d}(k,\protect\varphi )$ with $k\in %
\left[ -\protect\pi ,0\right] $; and (b) $\boldsymbol{d}(k,\protect%
\varphi )$ with $k\in \left[ 0,\protect\pi \right] $.}
\label{2sphere}
\end{figure}
This is because the trajectories of the vector $\boldsymbol{d}(k,\varphi )$
can be split into two equal disjoint topological spheres $\boldsymbol{d}%
(k,\varphi )$ with $k\in \left[ -\pi ,0\right] $, and $\boldsymbol{d}%
(k,\varphi )$ with $k\in \left[ 0,\pi \right] $ (see Fig. \ref{2sphere}).
Note that each of the topological spheres will contribute a Euler number 2 ($%
\chi =2(1-g)$, with $g=0$ for a topological sphere), and hence the two
topological spheres will contribute the Euler number $\chi =4$.

For the measurable consequence, Neupert et. al. have recently shown that the
quantum geometric tensor $Q_{\mu {\nu }}$ of the band can be measured by the
current noise spectrum of the band insulators. In a two-band Hamiltonian,
the current noise spectrum can be expressed by the $Q_{\mu {\nu }}$ of the
band as (see Eq. (13) in Ref. \cite{Neupert2})
\begin{equation}
S_{\mu {\nu }}(\omega )=-2\pi \omega ^{2}\int_{1\text{Bz}}\frac{d^{d}%
\boldsymbol{k}}{\Omega _{\text{Bz}}}\delta \left[ \omega -E_{+}(\boldsymbol{k%
})+E_{-}(\boldsymbol{k})\right] Q_{\mu {\nu }}(\boldsymbol{k}),
\end{equation}%
where $\Omega _{\text{BZ}}$ denotes the volume of the Brillouin zone, $d$
denotes the dimension of the crystal momentum space, and
\begin{equation}
S_{\mu {\nu }}(\omega ):=\int dt\ e^{-i\omega t}\langle {0}|J_{\mu }(0)J_{{%
\nu }}(t)\left\vert {0}\right\rangle
\end{equation}%
is the spectral function of the current-current correlation. As shown by
Marzari and Vanderbilt \cite{Marzari}, the integral of the trace of the
Riemannian metric over the Brillouin zone $\Omega _{I}=\int_{1\text{Bz}}%
\frac{d^{d}\boldsymbol{k}}{\Omega _{\text{Bz}}}\text{Tr}\mathcal{G}$ is a gauge
invariant measure of the delocalization or spread of the Wannier functions.
Here, we would like to point out that the Euler number of the band, in the
case of a 2D two-band Hamiltonian, can be reduced to a gauge invariant volume
of the Brillouin zone as measured according to the metric $\mathcal{G}$ (see
Eq. (\ref{eu2}) ) and this volume can be topological invariant in the
nontrivial topological phase.

\begin{figure}[h]
\begin{center}
\includegraphics[width=0.23\textwidth]{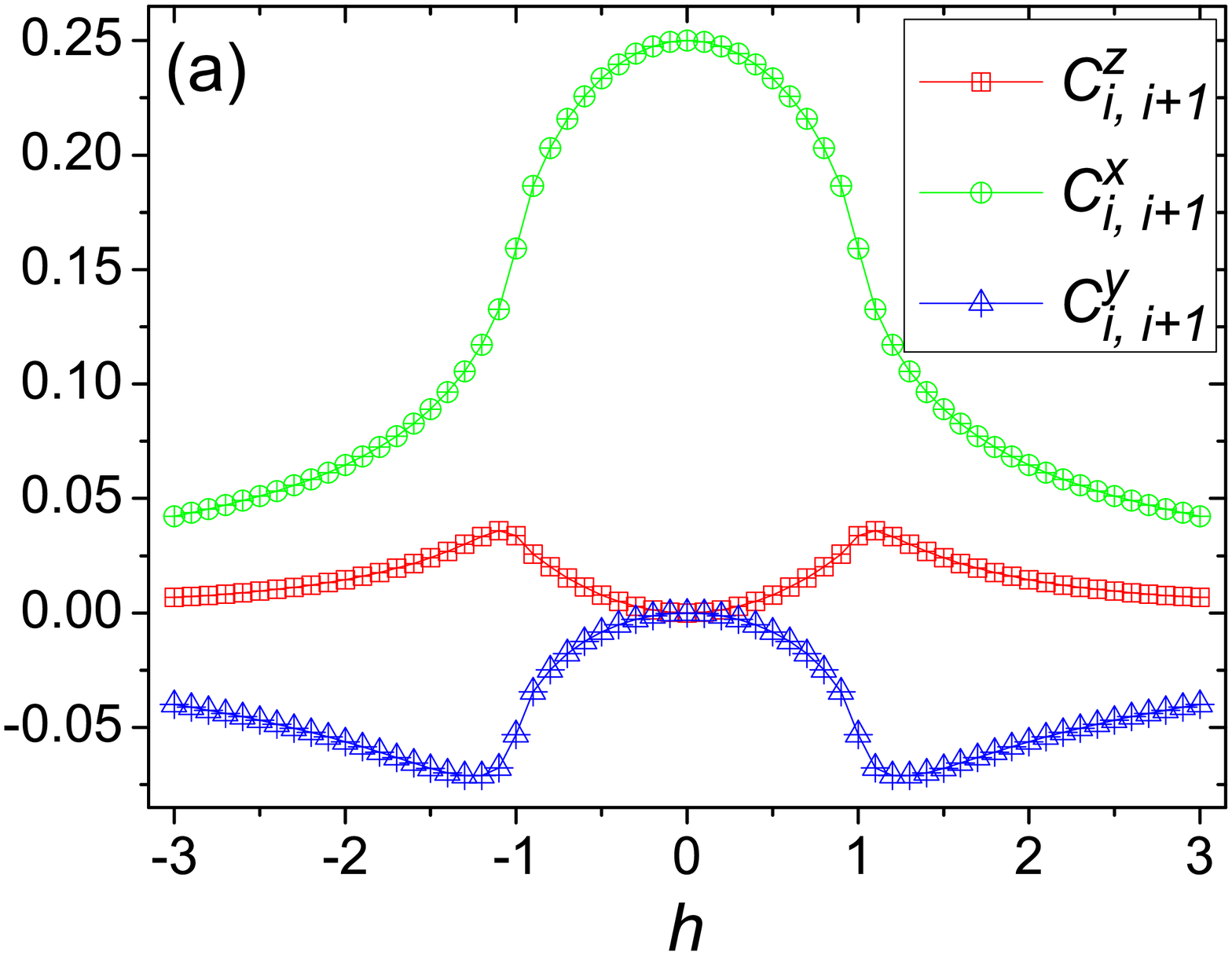} %
\includegraphics[width=0.23\textwidth]{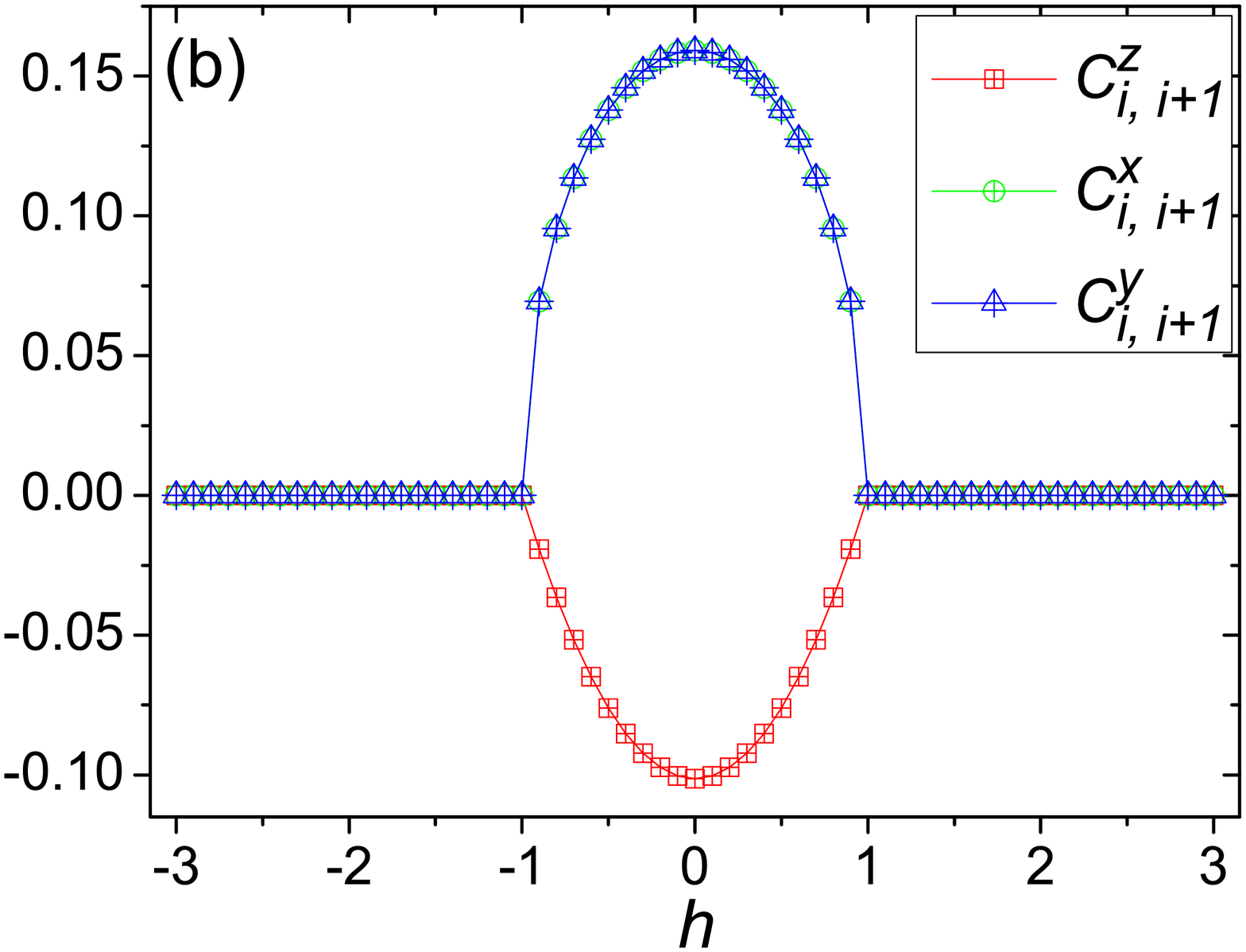} %
\includegraphics[width=0.23\textwidth]{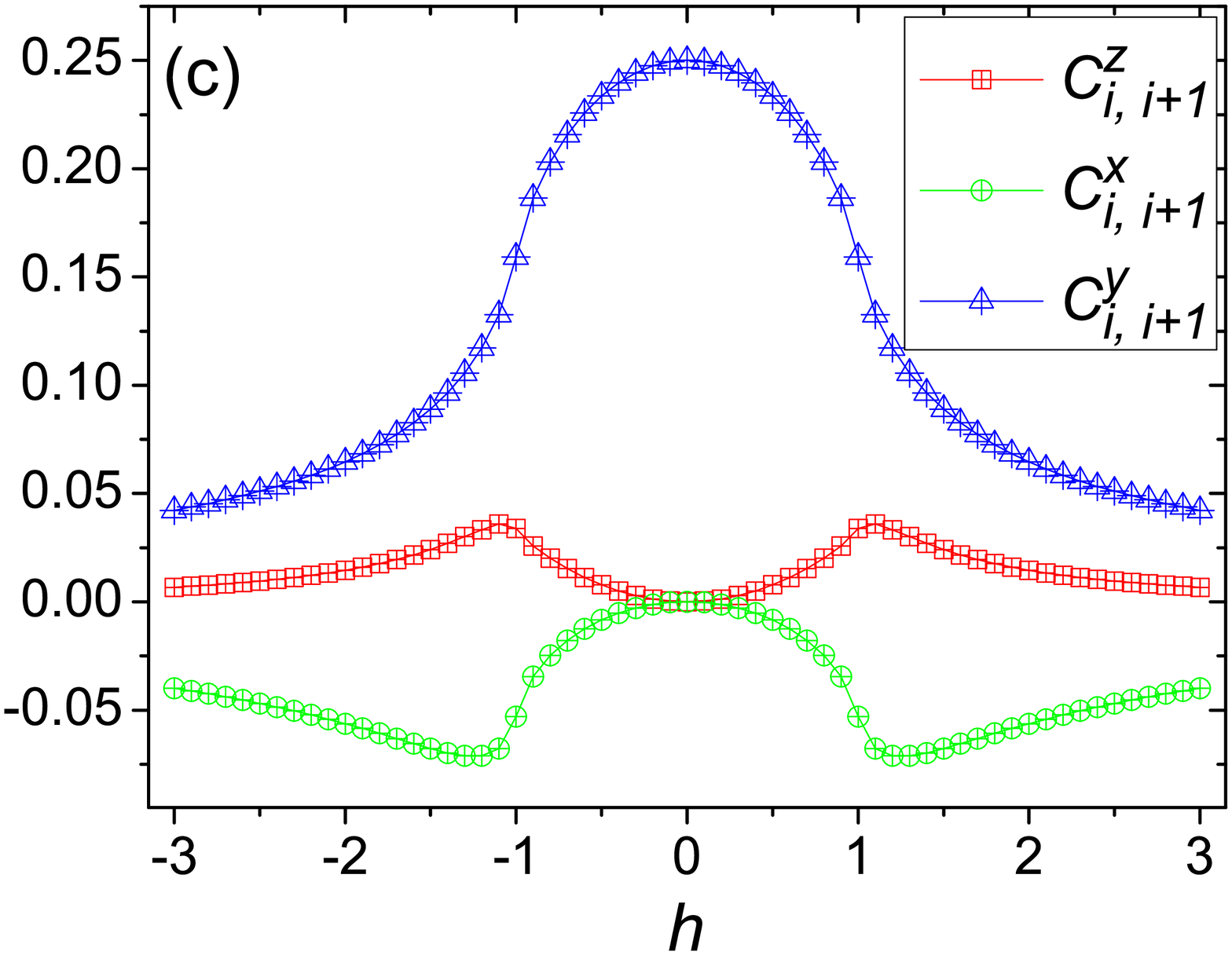} %
\includegraphics[width=0.23\textwidth]{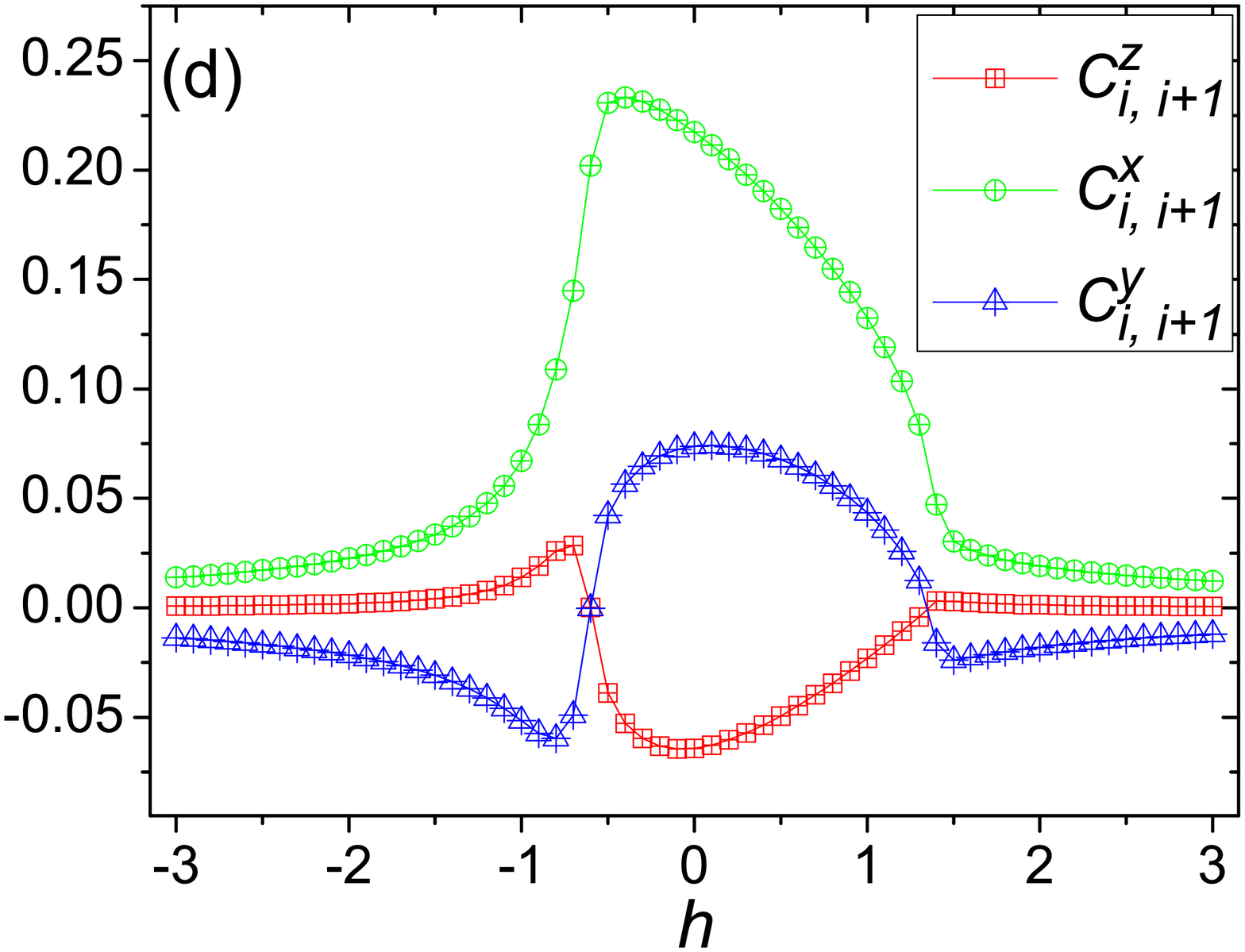} %
\includegraphics[width=0.23\textwidth]{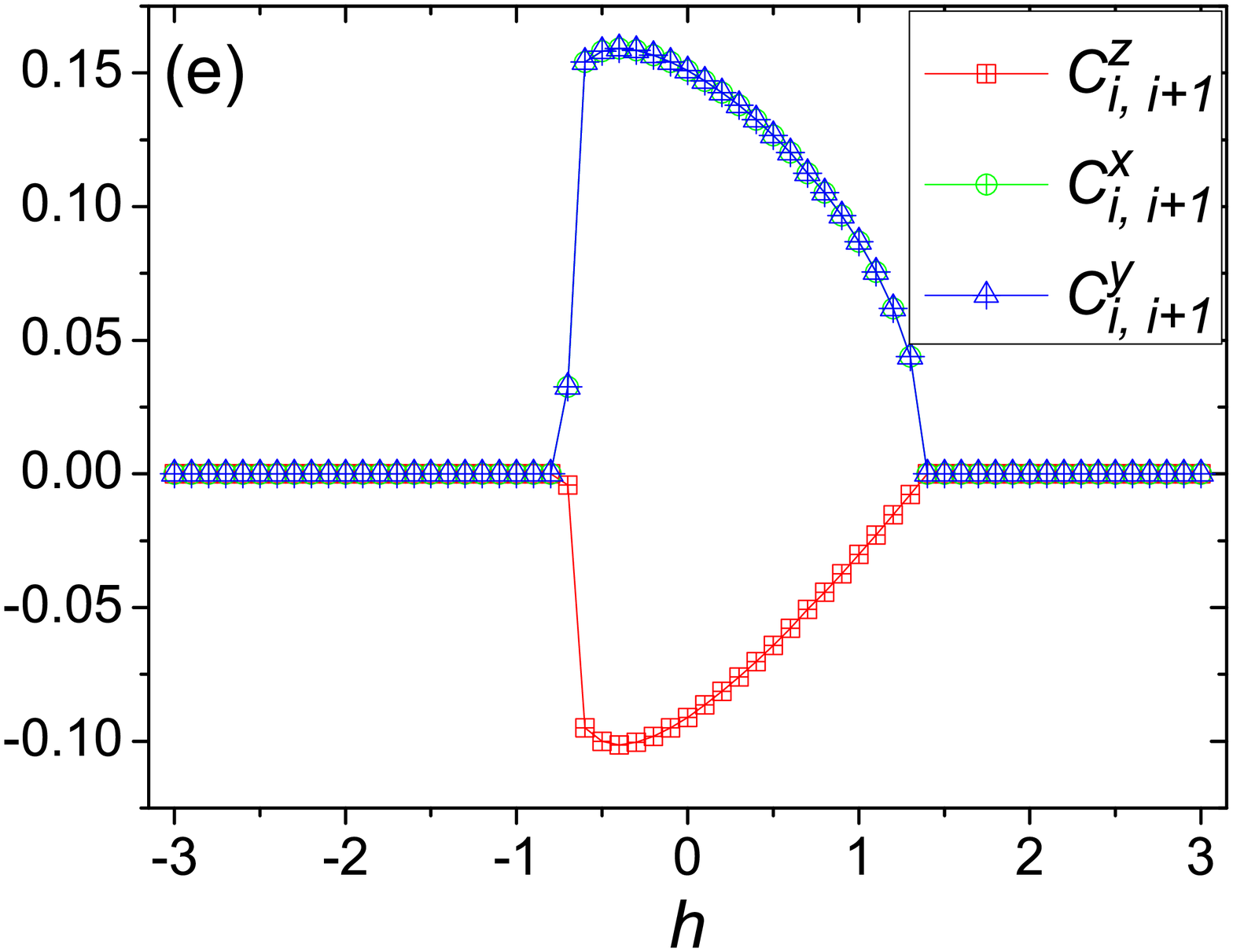} %
\includegraphics[width=0.23\textwidth]{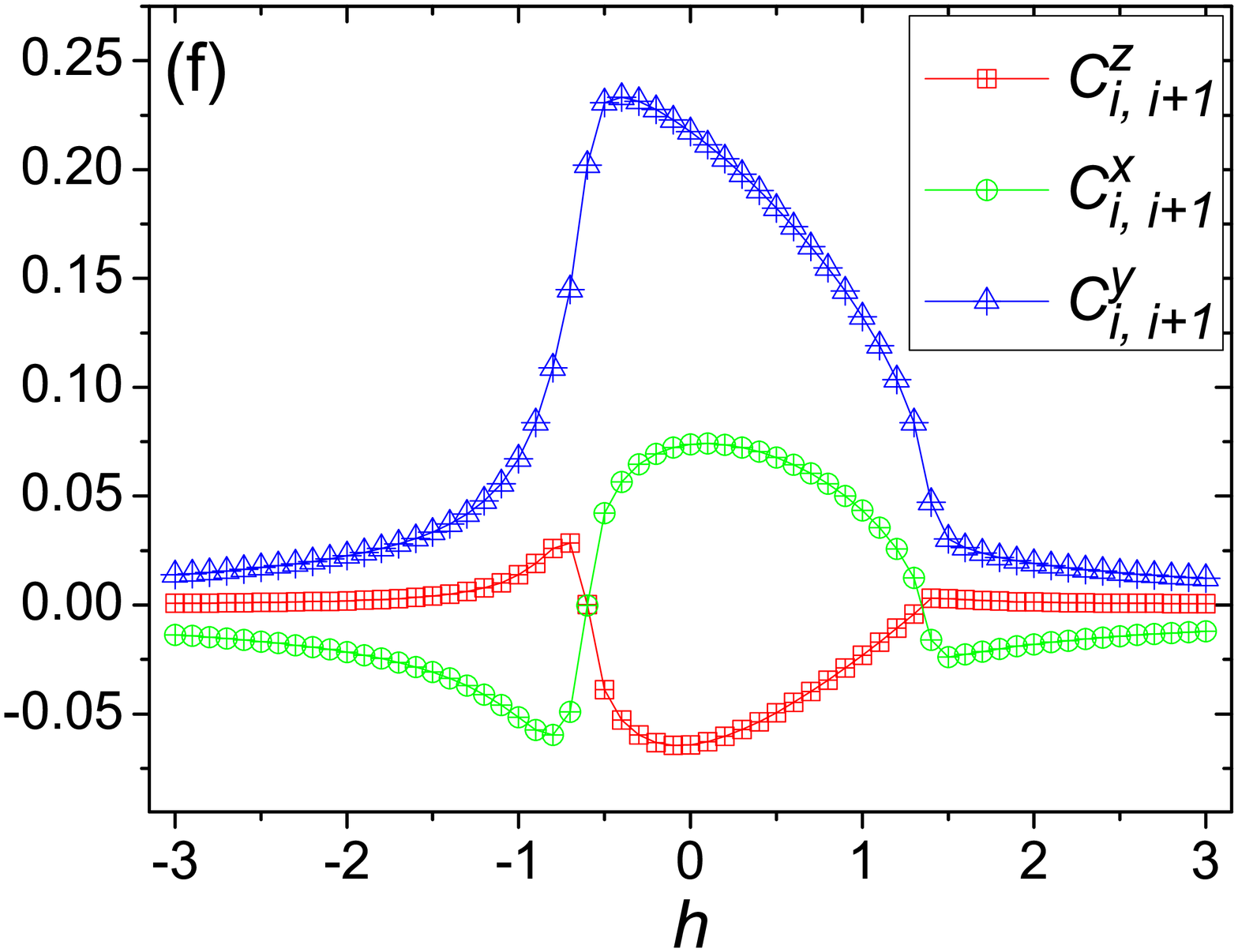}
\end{center}
\caption{(color online) The nearest neighbors spin-spin correlations $%
C_{i,i+1}^{x}$, $C_{i,i+1}^{y}$, and $C_{i,i+1}^{z}$ as functions of $h$,
with the fixed parameters (a): $\protect\gamma =1,\protect\delta =0$; (b): $%
\protect\gamma =0,\protect\delta =0$; (c): $\protect\gamma =-1,\protect%
\delta =0$; (d): $\protect\gamma =0.3,\protect\delta =0.4$; (e): $\protect%
\gamma =0,\protect\delta =0.4$; (f): $\protect\gamma =-0.3,\protect\delta %
=0.4$.}
\end{figure}

The spin-spin correlation functions can be derived by using the similar
method for the transverse field XY spin model (see Refs. \cite%
{ent2} and Ref. \cite{Suzuki}), and we can obtain%
\begin{eqnarray}
C_{i,i+r}^{x} &=&\langle S_{0}^{x}S_{r}^{x}\rangle -\langle S_{0}^{x}\rangle
\langle S_{r}^{x}\rangle  \notag \\
&=&\frac{1}{4}\left\vert
\begin{array}{cccc}
G_{-1} & G_{-2} & \cdot & G_{-r} \\
G_{0} & G_{-1} & \cdot & G_{-r+1} \\
\vdots & \vdots & \ddots & \vdots \\
G_{r-2} & G_{r-3} & \cdot & G_{-1}%
\end{array}%
\right\vert ,  \notag \\
C_{i,i+r}^{y} &=&\langle S_{0}^{y}S_{r}^{y}\rangle -\langle S_{0}^{y}\rangle
\langle S_{r}^{y}\rangle  \notag \\
&=&\frac{1}{4}\left\vert
\begin{array}{cccc}
G_{1} & G_{0} & \cdot & G_{-r+2} \\
G_{2} & G_{1} & \cdot & G_{-r+3} \\
\vdots & \vdots & \ddots & \vdots \\
G_{r} & G_{r-1} & \cdot & G_{1}%
\end{array}%
\right\vert ,  \notag \\
C_{i,i+r}^{z} &=&\langle S_{0}^{z}S_{r}^{z}\rangle -\langle S_{0}^{z}\rangle
\langle S_{r}^{z}\rangle =-\frac{1}{4}G_{r}G_{-r},  \label{spinspin}
\end{eqnarray}%
where the expectation values $\langle $ $\rangle $ are taken in the ground
state (at zero temperature) or in the canonical ensemble (at finite
temperature), and the Green function $G_{r}$ at finite temperature is given
by%
\begin{eqnarray}
G_{r}\left( \beta \right) &=&\frac{1}{\pi }\int_{0}^{\pi }\left( h+\cos
k-\delta \cos 2k\right) \cos kr\frac{\tanh \left( \frac{1}{2}\beta \Lambda
_{k}\right) }{\Lambda _{k}}  \notag \\
&&-\gamma \sin k\sin kr\frac{\tanh \left( \frac{1}{2}\beta \Lambda
_{k}\right) }{\Lambda _{k}}dk,  \label{Grfinite}
\end{eqnarray}%
where $\beta =k_{B}T$ and the energy spectrum $\Lambda _{k}=$ $\sqrt{\left(
h+\cos k-\delta \cos 2k\right) ^{2}+\gamma ^{2}\sin ^{2}k}$. Meanwhile, the
Green function $G_{r}$ at zero temperature can be obtained by setting $\tanh
\left( \beta \Lambda _{k}/2\right) =1$, that is,
\begin{equation}
G_{r}=\frac{1}{\pi }\int_{0}^{\pi }\frac{\left( h+\cos k-\delta \cos
2k\right) \cos kr-\gamma \sin k\sin kr}{\sqrt{\left( h+\cos k-\delta \cos
2k\right) ^{2}+\gamma ^{2}\sin ^{2}k}}dk.  \label{Gr}
\end{equation}%
The nearest neighbors spin-spin correlation functions at zero temperature as
functions of $h$ with different Hamiltonian parameters $\gamma $ and $\delta
$ have been shown in Fig. 10.

On the other hand, the Euler number of the system can be expressed as (see
Eq. (\ref{eu2}))
\begin{equation}
\chi =\frac{1}{2\pi }\int_{1\text{Bz}}4\sqrt{\mathcal{G}_{kk}\cdot \mathcal{G%
}_{\varphi \varphi }}d{k}d\varphi ,  \label{Euler}
\end{equation}%
where the Riemannian metric $\mathcal{G}_{kk}$ and $\mathcal{G}_{\varphi
\varphi }$ are given by Eq. (\ref{metrictensor}). Considering Eqs. (\ref%
{spinspin}), (\ref{Gr}), and (\ref{metrictensor}), we have%
\begin{equation}
\frac{1}{2\pi }\iint \sin k\sqrt{\mathcal{G}_{\varphi \varphi }}d{k}d{%
\varphi }=\frac{\pi }{4}\left( C_{i,i+1}^{x}-C_{i,i+1}^{y}\right) .
\end{equation}%
By a similar way, the metric component $\mathcal{G}_{kk}$ and the Euler
number can also be expressed in some combination of the spin-spin
correlation functions.

\begin{figure}[h]
\begin{center}
\includegraphics[width=0.23\textwidth]{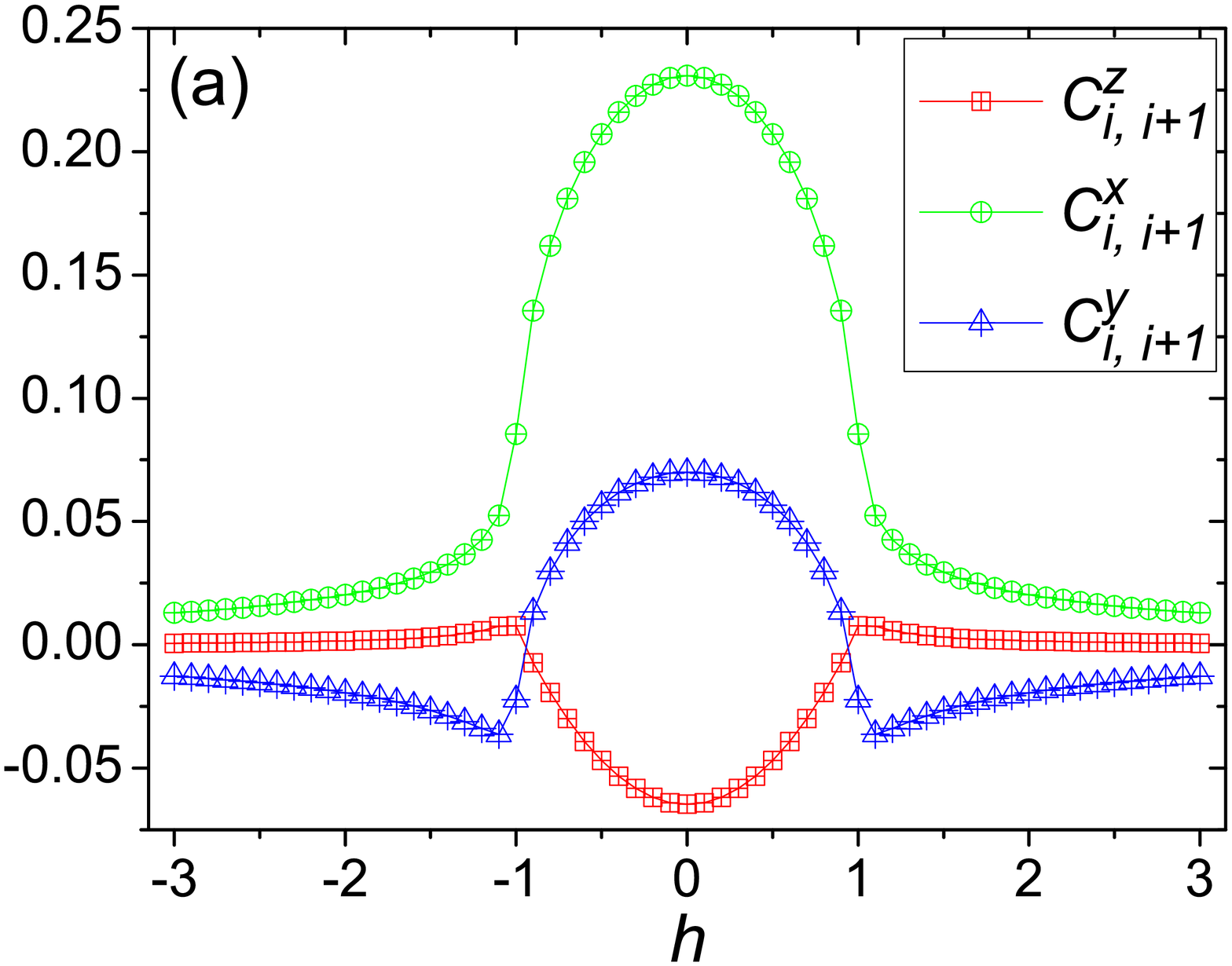} %
\includegraphics[width=0.23\textwidth]{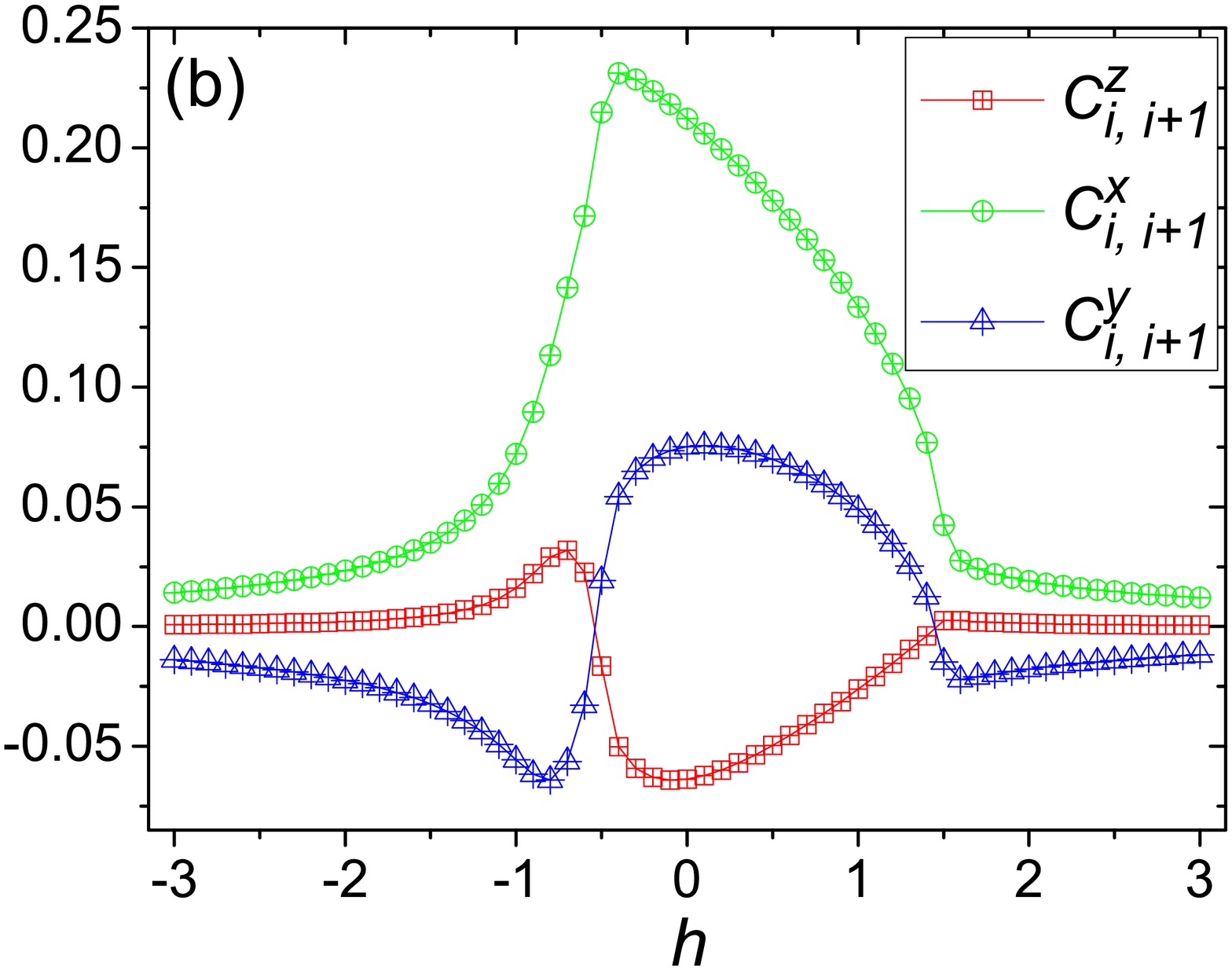} %
\includegraphics[width=0.23\textwidth]{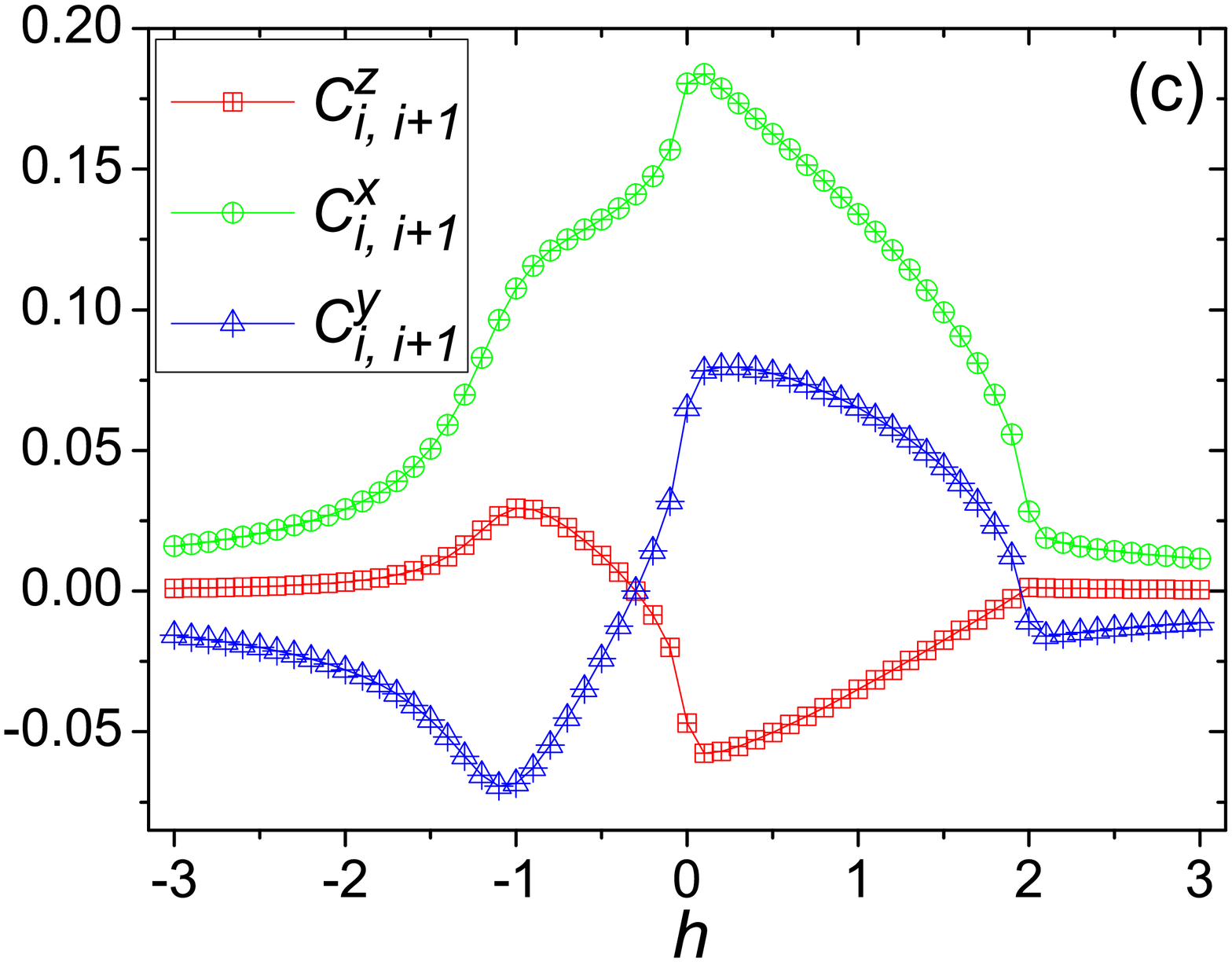} %
\includegraphics[width=0.23\textwidth]{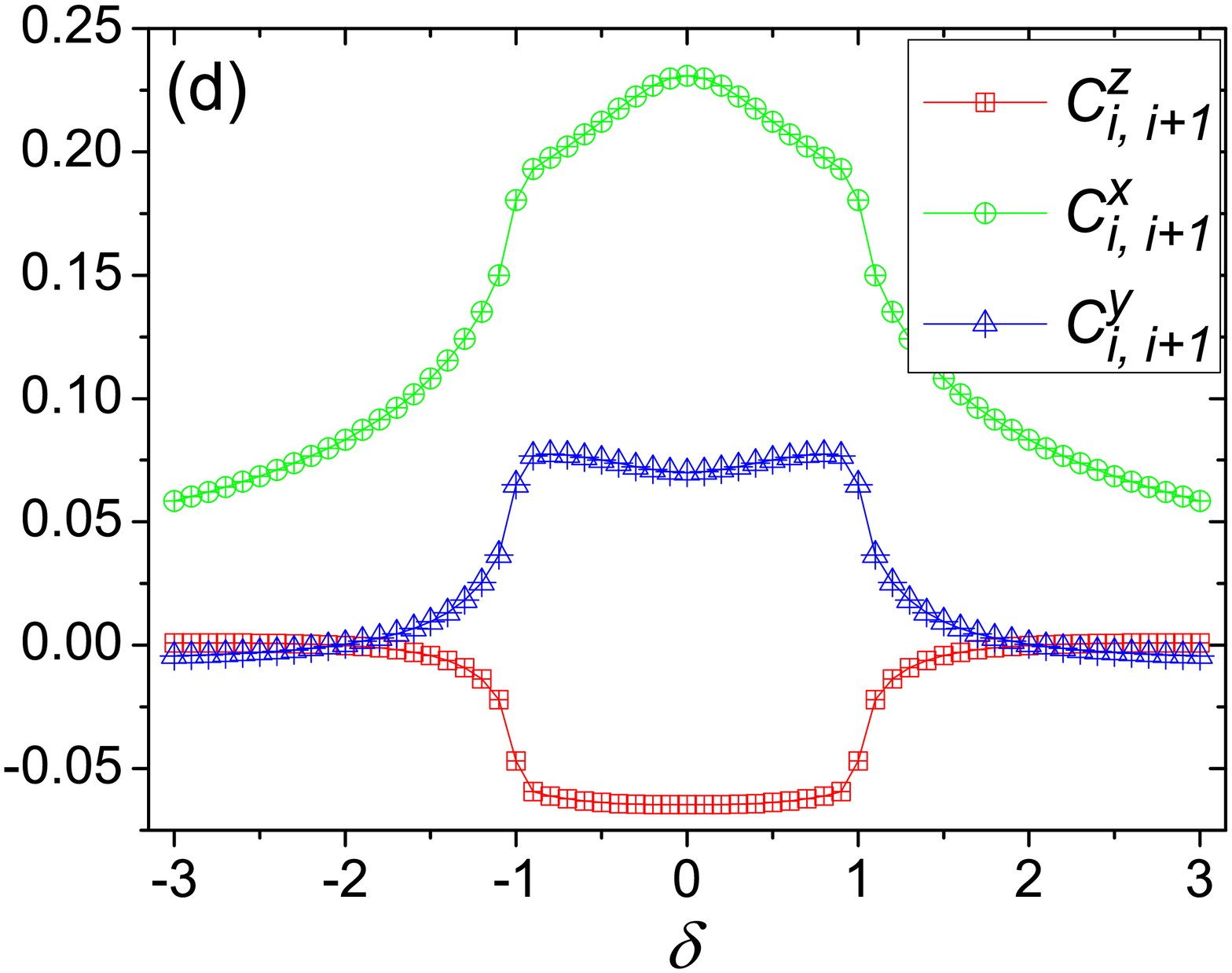} %
\includegraphics[width=0.23\textwidth]{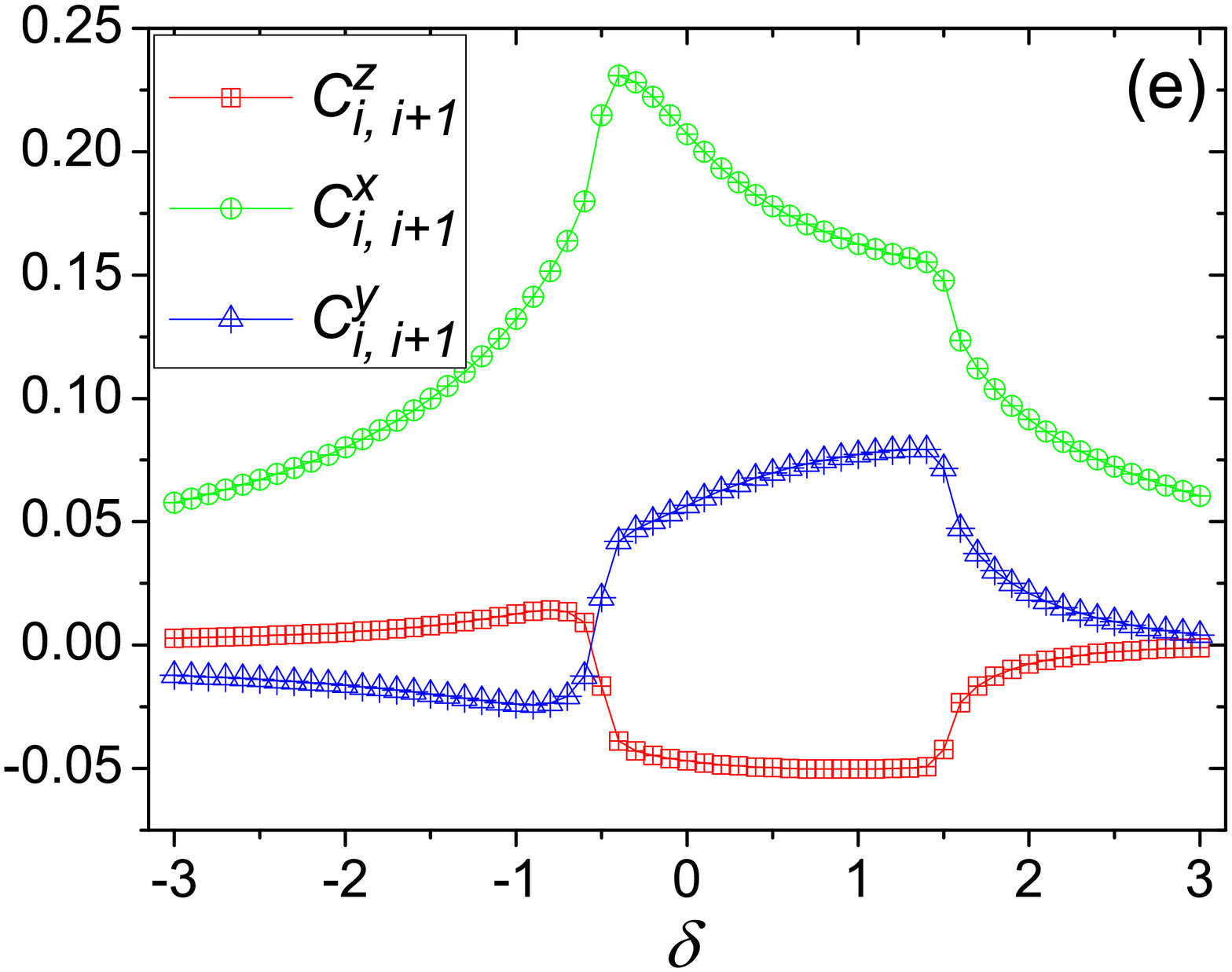} %
\includegraphics[width=0.23\textwidth]{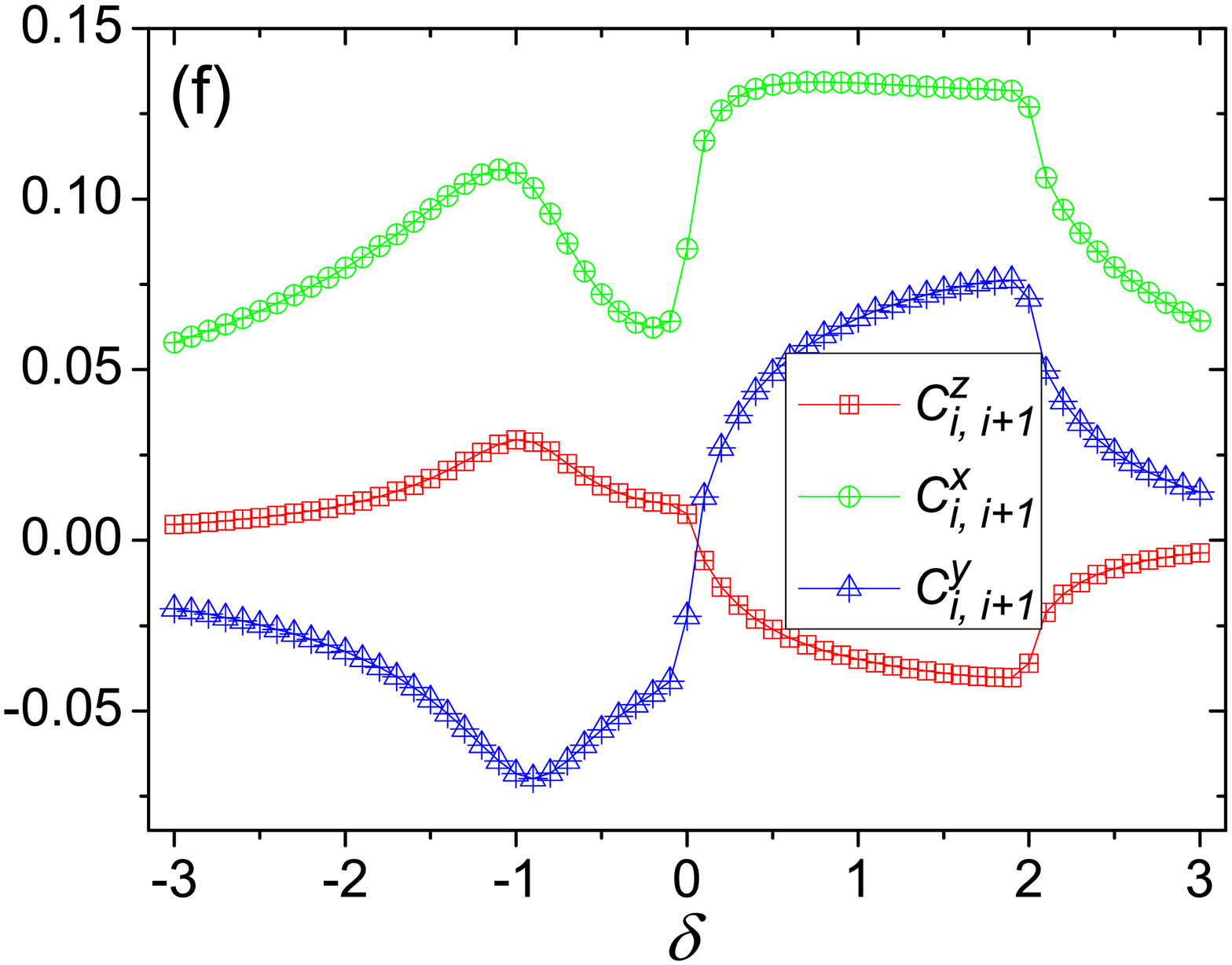}
\end{center}
\caption{(color online) The spin-spin correlations $C_{i,i+1}^{x}$, $%
C_{i,i+1}^{y}$, and $C_{i,i+1}^{z}$ as the function of $h$ with the fixed
parameters $\protect\gamma =0.3$, and (a): $\protect\delta =0$; (b): $%
\protect\delta =0.5$; (c): $\protect\delta =1$; The spin-spin correlations $%
C_{i,i+1}^{x}$, $C_{i,i+1}^{y}$, and $C_{i,i+1}^{z}$ as the function of $%
\protect\delta$ with the fixed parameters $\protect\gamma =0.3$, and (d): $%
h=0$; (e): $h=0.5$; (f): $h=1$.}
\end{figure}

In Figs. 11(a), 11(b), and 11(c), we show the correlation functions $%
C_{i,i+1}^{x}$, $C_{i,i+1}^{x}$, and $C_{i,i+1}^{z}$ as the functions of the
transverse field $h$ with the Hamiltonian parameters $\gamma =0.3$, and $%
\delta =0$, $0.5$, $1$. In Figs. 11(d), 11(e), and 11(f), we show the
correlation functions as the functions of the three-site spin coupling
coefficient $\delta $ with the Hamiltonian parameters $\gamma =0.3$, and $h=0
$, $0.5$, $1$. As shown in Fig. 11, the three-site spin coupling coefficient
$\delta $ will affect the behavior of the spin-spin correlation functions.
Meanwhile, the critical points of the system will be moved to $h=\delta \pm 1$
where the energy gap will be closed, and this can be witnessed by the Euler
number of the band.

\section{Conclusions}

In summary, we study the Euler number index of the Bloch band in a
transverse field XY spin-1/2 chain with multi-site spin couplings. This
approach is based on the topological characterization from the Gauss-Bonnet
theorem on a 2D closed Bloch states manifold in the first Brillouin zone,
where the Riemannian structure of the Bloch band is established by the
geometric tensor in the crystal momentum space. For a local geometric
witness to the quantum phase transitions, we introduce the cyclic quantum
distance of the Bloch band and show the Riemannian metric on the Bloch
states manifold can be relate to a corresponding ground-state quantum
distance in the parameter space. Finally, we derive the Euler characteristic
number of the Bloch band analytically via the Gauss-Bonnet theorem on the 2D
Bloch states manifold in the first Brillouin zone. We show that the
ferromagnetic-paramagnetic quantum phase transition in this model is
topologically different in the Bloch band's Euler number index. We also give a
general formula of the Euler number for the 1D or 2D two-band systems, which
reveals its essential relation to the first Chern number of the band
insulators.

\section{Acknowledgments}

The author would like to thank Han Zhang for helpful discussions and thank the referee for the constructive comments
which helped to improve the manuscript. This work was supported by the Special Foundation for Theoretical Physics Research of NSFC under Grant No. 11347131 and NSFC under Grant No. 11404023.


\begin{thebibliography}{99}
\bibitem{note1} Note the Riemannain metric of the Bloch states $u_{\pm
}\left( k,\varphi \right) $ is given by $\mathcal{G}_{k\varphi }^{\pm }=%
\frac{1}{2}\langle \partial _{k}u_{\pm }|\partial _{\varphi }u_{\pm }\rangle
+\frac{1}{2}\langle \partial _{\varphi }u_{\pm }|\partial _{k}u_{\pm
}\rangle -\langle \partial _{k}u_{\pm }|u_{\pm }\rangle \langle u_{\pm
}|\partial _{\varphi }u_{\pm }\rangle $, and it can be verified that here $%
\mathcal{G}_{k\varphi }^{+}=\mathcal{G}_{k\varphi }^{-}$, so we can denote
the metric as $\mathcal{G}_{k\varphi }$ conveniently.
\end{thebibliography}
\end{document}